\documentclass[fleqn,usenatbib]{mnras}

\usepackage{newtxtext,newtxmath}
\usepackage{soul} 
\usepackage{xcolor}
\usepackage{float}
\usepackage{capt-of}
\usepackage{tabularx}
\usepackage{booktabs}
\usepackage[T1]{fontenc}
\usepackage[version=4]{mhchem}
\usepackage{soul}
\DeclareRobustCommand{\VAN}[3]{#2}
\let\VANthebibliography\thebibliography
\def\thebibliography{\DeclareRobustCommand{\VAN}[3]{##3}\VANthebibliography}

\usepackage{graphicx}	
\usepackage{amsmath}	
\usepackage{flushend}
\usepackage{caption}
\usepackage{comment}
\usepackage{placeins}
\usepackage{bm}

\title[Ethanolamine–water–methanol ices]{Electron irradiation and temperature-programmed desorption of layered ethanolamine–water–methanol ices}

\author[Tara L. Stoib et al.]{Tara L. Stoib,$^{1,2}$\thanks{Present address: Zernike Institute for Advanced Materials, University of Groningen, Nijenborgh 4, 9747 AG Groningen, The Netherlands} Alejandro Guerrero-Caicedo,$^{3,4,5}$ Duncan V. Mifsud,$^6$ Péter Herczku,$^6$  Gergő Lakatos,$^{6,7}$ \newauthor Zuzana Kaňuchová,$^{6,8}$ Zoltán Juhász,$^6$ Béla Sulik,$^6$ Nigel J. Mason,$^{6,9}$ Sergio Ioppolo,$^{10}$ Paola Caselli,$^1$ \newauthor Barbara Michela Giuliano,$^1$ Karl E. Duderstadt,$^{2,11}$ Felipe Fantuzzi,$^{3,6,12}$ and Heidy M. Quitián-Lara$^{1,3,9}$\thanks{Email: heidyql@mpe.mpg.de}
\\
$^{1}$Center for Astrochemical Studies, Max-Planck-Institut für Extraterrestrische Physik, Garching, Germany\\
$^{2}$Department of Bioscience, Technical University of Munich, Garching, Germany\\
$^{3}$Síntesis y Mecanismos de Reacción en Química Orgánica, Departamento de Química, Universidad del Valle, Cali, Colombia\\
$^{4}$Semillero de investigación en Astroquímica y Astrobiología (AstroLUCA), Departamento de Química, Universidad del Valle, Cali, Colombia\\
$^{5}$Faculty of Health Sciences, Universidad Libre, Cali, Colombia\\
$^{6}$HUN-REN Institute for Nuclear Research, Debrecen, Hungary\\
$^{7}$Institute of Chemistry, University of Debrecen, Debrecen, Hungary\\
$^{8}$Astronomical Institute, Slovak Academy of Sciences, Tatranská Lomnica SK-059 60, Slovakia\\
$^{9}$Centre for Astrophysics and Planetary Science, School of Engineering, Mathematics, and Physics, Canterbury, United Kingdom\\
$^{10}$Center for Interstellar Catalysis, Department of Physics and Astronomy, Aarhus University, Aarhus, Denmark\\
$^{11}$Structure and Dynamics of Molecular Machines, Max Planck Institute of Biochemistry, Martinsried, Germany\\
$^{12}$Supramolecular, Interfacial and Synthetic Chemistry, School of Natural Sciences, University of Kent, Canterbury, United Kingdom\\}

\date{Accepted XXX. Received YYY; in original form ZZZ}

\pubyear{\the\year{}}

\begin{document}
\label{firstpage}
\pagerange{\pageref{firstpage}--\pageref{lastpage}}
\maketitle

\begin{abstract}
Ethanolamine (EtA) is a recently detected interstellar complex organic molecule and a potential precursor to more functionalised N-bearing species. We investigate the electron-driven chemistry of layered EtA--H$_2$O--CH$_3$OH ices deposited at 20~K, combining 2~keV electron irradiation with subsequent temperature-programmed desorption (TPD) while monitoring the solid phase by mid-infrared spectroscopy. These controlled irradiation conditions represent a simplified laboratory analogue of the structural and energetic conditions within interstellar grain mantles. Volatile products released during warming were followed by quadrupole mass spectrometry, and the final residue was analysed by \textit{ex situ} electrospray ionisation mass spectrometry (ESI-MS). Irradiation produces abundant radiolysis products typical of processed methanol- and water-rich ices, including CO, CO$_2$, H$_2$CO, H$_2$O$_2$, NH$_3$, and OCN$^{-}$. Bands consistent with formic acid and polyoxymethylene-like material are also observed, indicating that carbonyl chemistry and formaldehyde-driven oligomerisation can occur alongside fragmentation in the layered system. Comparison with a choline chloride--H$_2$O reference reveals the emergence of features compatible with methylated EtA derivatives and/or choline-like species, while ESI-MS shows higher-$m/z$ mass families in the residue that are not reproduced in the substrate/background blank. However, spectral congestion and matrix effects preclude an unambiguous assignment of choline. Our results show that energetic processing of EtA-containing ices can generate a diverse inventory of O- and N-bearing products, providing laboratory constraints on radiation chemistry in cold, irradiated grain mantles under idealised conditions.

\end{abstract}




\section{Introduction}

\begin{figure}
\centering
\includegraphics[width=\columnwidth]{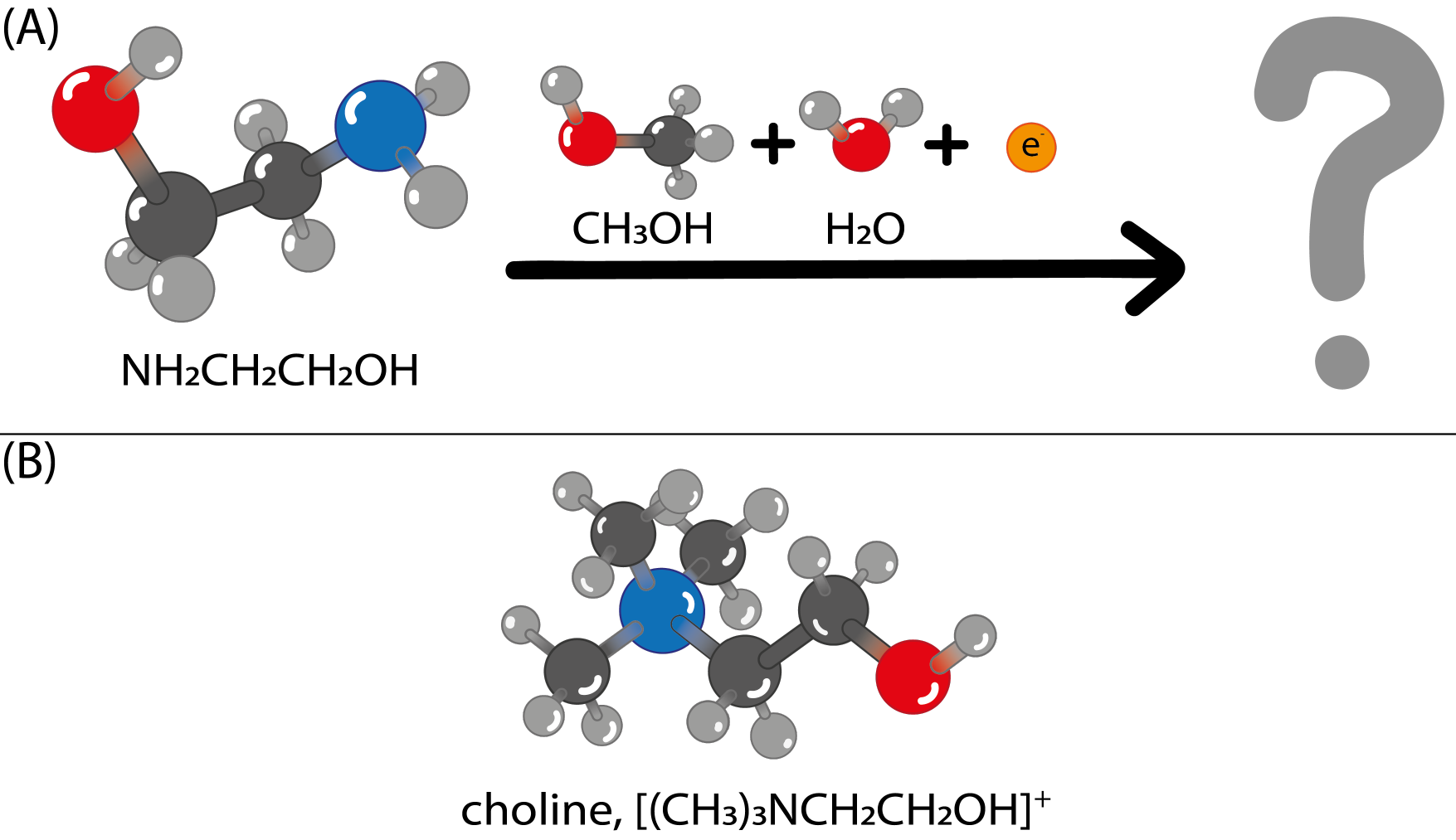}
\caption{Schematic representation of the chemical concept explored in this work. 
(A) Ethanolamine (EtA, NH$_2$CH$_2$CH$_2$OH), methanol (CH$_3$OH), and water (H$_2$O) are subjected to electron irradiation, illustrating the starting molecular components and the formation of an unresolved product inventory after energetic processing. The question mark denotes the chemically diverse products investigated by FTIR, TPD-QMS, and ESI-MS. 
(B) Structure of the choline cation, [(CH$_3$)$_3$NCH$_2$CH$_2$OH]$^{+}$, included as a structurally related reference for assessing methylated EtA-derived and/or choline-like spectral features.}
\label{fig:Overview_exp_reac}
\end{figure}

Observations over recent years have shown that interstellar molecules are widespread and can reach significant complexity \citep{Xue2025}, with a growing diversity of chemical species detected across a wide range of astrophysical environments \citep{Ehrenfreund2000}. This chemical richness is central to many astrophysical and chemical processes and has driven sustained interest in complex organic molecules (COMs), commonly defined as astrochemically relevant organic species containing six or more atoms \citep{HerbstCOIM2009,fulvioAstrochemicalPathwaysComplex2021}. COMs have been observed in a broad range of astrophysical environments, including interstellar clouds \citep{ZuckermanFormicAcid1971,Bacmann2012_PRESTELLAR_COMs,RodriguezAlmeidaDetection2021}, pre-stellar cores \citep{vastel_origin_2014,jimenez-serra_spatial_2016,Scibelli2021_L1521E_COMs,Lin2026}, protostars \citep{Lefloch2018,Quitian-Lara2023,Taniguchi2024}, protoplanetary discs \citep{Oberg2015_MWC480_CH3CN,Walsh2016_TWHydrae_CH3OH}, comets \citep{Biver2015_Lovejoy_Ethanol_Glycolaldehyde,Hanni2025}, and extragalactic sources \citep{Martin2021}. Nevertheless, the majority of COM detections have been reported in molecular clouds, with a significant fraction of this molecular complexity concentrated in the Central Molecular Zone (CMZ) near the Galactic Centre (GC; \citealt{jimenez2025chemistry}), particularly in sources such as Sagittarius B2 (Sgr B2; \citealt{belloche2013complex,Xue2025}). The extreme physical conditions in the CMZ, including elevated temperatures, high densities, and intense irradiation fields, make this region a natural laboratory for investigating complex chemical networks \citep{Armillotta2020,Bryant2021,Su2024}. Among the COMs identified to date is ethanolamine (EtA, NH$_2$CH$_2$CH$_2$OH, Figure~\ref{fig:Overview_exp_reac}(A)), which was detected by \citet{rivillaDiscoverySpaceEthanolamine2021} in the molecular cloud G+0.693$-$0.027 towards the GC. The presence of both an alcohol and an amine functional group makes EtA of particular astrochemical and astrobiological interest, as it may provide a link between relatively simple interstellar molecules and more complex prebiotic species \citep{Bakovic2007,sladkovaDestructionAminoAlcohols2014,quitian-laraPhotodissociationEthanolamineInterstellar2025}. Beyond the gas-phase detection in G+0.693$-$0.027, ethanolamine is also relevant in a broader prebiotic context: meteoritic analyses have reported amino-acid inventories in carbonaceous material \citep{GLAVIN2010}, and UV processing of interstellar ice analogues has been shown to produce racemic amino acids \citep{Bernstein2002, MunozCaro2002}, motivating continued efforts to characterise how N- and O-bearing organics persist and transform under energetic processing \citep{Mason2026a}. Recent astrochemical reviews have highlighted simple amphiphiles as plausible ancestors of membrane-forming molecules and discussed how interstellar chemistry may have contributed to early terrestrial inventories, with lipids among the most robust candidate biomarkers \citep{Bockova2024}.

EtA is abundant in biological systems, where it serves as the head group of phosphatidylethanolamine (PE), a major phospholipid component of cellular membranes that plays a key role in membrane structure, curvature, and dynamics \citep{yeagleLipidsBiologicalMembranes2016}. EtA is also chemically linked to choline (Figure~\ref{fig:Overview_exp_reac}(B); \citealt{Kewith1976}), phosphatidylcholine (PC), and glycine \citep{zhangPrebioticSynthesisGlycine2017}, and functions as a growth factor that stimulates the rapid proliferation of mammalian cells in culture \citep{patelEthanolaminePhosphatidylethanolaminePartners2017}. In particular, stepwise N-methylation of EtA is a recognised route to choline formation \citep{aleksandrovaRingOpeningGlycerol2023}. Consequently, the detection of EtA in astrophysical environments raises fundamental questions regarding its formation under pre-stellar conditions, its stability under irradiation, and its potential role as a node connecting nitrogen chemistry to more complex prebiotic species and proto-amphiphilic inventories \citep{Bockova2024}.

EtA has been the focus of numerous studies investigating its formation under interstellar conditions \citep{rivillaDiscoverySpaceEthanolamine2021,ramachandranExperimentalComputationalStudy2024}, its fragmentation pathways and photostability in astrophysical environments \citep{zhangSystematicIRVUV2024,quitian-laraPhotodissociationEthanolamineInterstellar2025,Suhasaria2025}, and its potential role as a precursor molecule in early Earth chemistry \citep{zhangPrebioticSynthesisGlycine2017,aleksandrovaRingOpeningGlycerol2023}. Possible ice-phase formation routes for EtA have been proposed by \citet{rivillaDiscoverySpaceEthanolamine2021} and \citet{ramachandranExperimentalComputationalStudy2024}. These include sequential hydrogenation of iminoethenone (HNCCO), as well as hydrogenation of methanimine (CH$_2$NH) to form aminomethyl-type intermediates, followed by non-diffusive radical--radical reaction with methanol-derived CH$_2$OH radicals. These ideas are reinforced by recent astrochemical modelling, which identifies aminoketene as an efficient grain-surface intermediate and predicts that subsequent hydrogenation can yield EtA with appreciable solid-phase abundances, while shocks may favour its release and survival in the gas phase \citep{Willis2025}. Complementary chemical-network and abundance simulations of the molecular cloud G+0.693$-$0.027 likewise emphasise shock-driven desorption and cosmic-ray-induced UV fields, and predict additional related species as potential observational targets \citep{Zhao2025}. From a broader physical-chemistry perspective, electron-impact ionisation cross sections for EtA and related prebiotic molecules have recently been computed up to keV energies, providing quantitative inputs for modelling electron-driven chemistry in irradiated regions \citep{Chakraborty2024a}.

In the laboratory, \citet{zhangSystematicIRVUV2024} investigated the fragmentation of pure EtA ice and EtA/H$_2$O ice mixtures under 1~keV electron bombardment, observing phase transitions during controlled warm-up experiments. \citet{Suhasaria2025} reported the formation of ethylene glycol and serine following Ly$\alpha$ irradiation of EtA ice and subsequent thermal processing. More recently, experiments combining temperature-programmed desorption and mass spectrometry examined EtA deposited on nanometric amorphous olivine grains at 10~K, showing that the dust substrate can strongly modify desorption profiles (trapping EtA to high temperatures) and that UV irradiation can yield a range of products formed directly on the grain surface \citep{Biancalani2024a}. In parallel, computational studies have broadened the astrochemical perspective on EtA itself: investigations of electron-driven fragmentation pathways have proposed high-energy routes that transform saturated COMs into unsaturated products with extended $\pi$ networks under cosmic-ray and shock-related processing \citep{Londono-Restrepo2025a}, while systematic surveys of the C$_2$H$_7$NO isomer in space have identified additional energetically low-lying candidates and provided spectroscopic parameters to guide future searches beyond EtA \citep{Noriega2025}.

Previous irradiation studies of pure EtA and EtA/H$_2$O ice analogues suggest that the solid-state chemistry is largely dominated by decomposition and dehydrogenation processes with the possible formation of several N-bearing species, but without clear evidence for products larger or chemically more complex than EtA itself \citep{zhangSystematicIRVUV2024,suhasariaInfraredSpectraSolidstate2024,Suhasaria2025}. This limitation motivates exploring increased ice complexity as a route to new reaction channels, particularly in the context of solid-phase astrochemistry on dust-grain ice mantles \citep{Dickers2025,Mason2026b}. Accordingly, introducing methanol (CH$_3$OH) provides a chemically richer and astrophysically relevant ice environment \citep{jimenez-serra_spatial_2016,mcclure_ice_2023,Vyjidak2026} while also offering access to additional reactive fragments, including CH$_3$-bearing intermediates, under energetic processing \citep{schmidtElectronInducedProcessingMethanol2021}. This approach is also chemically relevant to the environment of G+0.693$-$0.027, where energetic processing and shocks appear to play an important role in the release and evolution of COMs \citep{Willis2025,Zhao2025}. The high column density of EtA reported in the molecular cloud G+0.693$-$0.027, $N=(1.51\pm0.07)\times10^{13}$~cm$^{-2}$, suggests efficient formation and/or survival in space \citep{rivillaDiscoverySpaceEthanolamine2021}. Furthermore, recent studies have demonstrated that EtA is relatively stable under astrophysical conditions, with estimated half-lives of $3.6\times10^{7}$~yr in cold dense clouds and $6.8\times10^{8}$~yr in Kuiper Belt Objects, indicating that EtA can persist when embedded in icy Solar System bodies \citep{zhangSystematicIRVUV2024}.

In this work, we investigate the irradiation-driven chemistry of EtA in layered H$_2$O- and CH$_3$OH-containing interstellar ice analogues to assess whether increasing ice complexity can promote molecular growth beyond simple fragmentation and to evaluate the potential role of EtA in the solid-state evolution of complex organic material in cold irradiated environments. Building on previous irradiation studies of EtA- and CH$_3$OH/H$_2$O-containing ices \citep{mifsudElectronIrradiationThermal2021,schmidtElectronInducedProcessingMethanol2021,zhangSystematicIRVUV2024,suhasariaInfraredSpectraSolidstate2024,Suhasaria2025,Stoib2026}, the present work extends this approach by combining infrared monitoring during irradiation and thermal desorption with complementary mass-spectrometric analysis of the final thermally processed residue. This allows volatile products, solid-phase spectral evolution, and less volatile residue components to be assessed together. Layered ices provide a useful experimental model because interstellar grain mantles can be chemically stratified as a consequence of sequential accretion, thermal processing, and energetic irradiation. Although the exact EtA--H$_2$O--CH$_3$OH geometry used here should be regarded as an idealised stratified ice rather than a direct reproduction of a specific interstellar mantle, it allows the effect of H$_2$O- and CH$_3$OH-rich material on EtA processing to be probed under controlled conditions \citep{Muller2022}. A direct comparison with fully mixed EtA--H$_2$O--CH$_3$OH ices is beyond the scope of the present work, and such experiments will be required to determine how ice morphology affects the relative efficiencies of fragmentation, radical recombination, and molecular growth.

\section{Methods}
\label{sec:methods}
\subsection{Experimental setup, ice preparation, and irradiation}
\label{sec:setup}

The experiments presented in this study were conducted at the HUN-REN Institute for Nuclear Research using the Ice Chamber for Astrophysics-Astrochemistry (ICA). The experimental setup has been described in detail elsewhere \citep{Herczku2021,mifsudElectronIrradiationThermal2021,zhangSystematicIRVUV2024}.

The ICA is an ultrahigh-vacuum chamber operating at a base pressure of approximately $10^{-9}$~mbar, maintained using a scroll pump coupled to a turbomolecular pump. At the centre of the chamber is a gold-coated oxygen-free high-conductivity copper sample holder hosting a series of infrared-transparent zinc selenide (ZnSe) substrates onto which astrophysical ice analogues may be deposited. The sample holder and deposition substrates can be cooled to a minimum temperature of 20~K through thermal contact with the cold finger of a closed-cycle helium cryostat, while the temperature can be controlled over the 20--310~K range using resistive heaters and a temperature control unit. Astrophysical ice analogues were prepared on the deposition substrates by dosing gases and vapours into the chamber through a fine needle valve, which then condensed into the solid phase upon contact with the cold substrates. When depositing vapours sourced from liquid samples, e.g. H$_2$O or CH$_3$OH, the liquid samples were purged of dissolved gases \textit{via} multiple freeze-pump-thaw cycles before the vapours were admitted into the main chamber.

Mid-infrared characterisation of the prepared ice analogues was performed \textit{in situ} using a Bruker V70v Fourier-transform infrared spectrometer and an external mercury-cadmium-telluride detector, which was used to acquire transmission spectra over a spectral range of $4000-650$~cm$^{-1}$ and at a spectral resolution of 0.5~cm$^{-1}$. Additionally, a quadrupole mass spectrometer (QMS; Pfeiffer QME200) was used for qualitative monitoring of gas-phase species and ion fragments up to $m/z=200$, with a mass resolution of 1~u.

In the present experiments, a layered EtA--H$_2$O--CH$_3$OH ice was prepared rather than a fully mixed ice. To enable the analysis of each deposited layer individually, mid-infrared spectra were recorded after the deposition of each component. The spectrum of each layer was then obtained by subtracting the spectrum recorded before its deposition from the spectrum recorded afterwards. Following this protocol, EtA was first deposited directly onto the substrate at 20~K. H$_2$O and CH$_3$OH were then introduced sequentially at the same temperature by background deposition, forming the second and third layers, respectively. The layered composition of the ice was selected to provide greater control over the amount of each deposited component and to allow the spectral contribution of each layer to be characterised separately. 

Following its preparation, the multilayer ice sample was irradiated for sixty minutes using a 2~keV electron beam supplied by a Kimball ELG-2A gun affixed to one of the side ports of the chamber. The flux of the electron beam was $3.2 \times 10^{13}$ electrons cm$^{-2}$ s$^{-1}$, corresponding to a total fluence of $1.15 \times 10^{17}$ electrons cm$^{-2}$ over the irradiation period. The projectile electrons impacted the target ice at an angle of 36$^\circ$ to the surface normal. The 2~keV electron beam was selected as a controlled laboratory source of electron-induced radiolysis in the layered ice. Although this energy does not reproduce the full energy distribution of galactic cosmic rays, keV electron irradiation has been widely used to investigate energetic processing of astrochemical ice analogues, including CH$_3$OH-, H$_2$O-, N$_2$O-, CH$_3$CN-, and sulphur-bearing ices \citep{Barnett2012,Mason2014,Ribeiro2015,Mifsud2022,mifsudElectronIrradiationThermal2021}\. Its relevance is further supported by the fact that energetic cosmic-ray particles interacting with molecular ices transfer a substantial fraction of their energy to the electronic system of the target molecules, producing cascades of secondary electrons with energies extending up to the keV range \citep{Bennett2007,arumainayagamExtraterrestrialPrebioticMolecules2019}. Importantly, keV electrons can produce electronic linear energy-transfer values of the same order of magnitude as those associated with MeV cosmic-ray particles in molecular ices \citep{Bennett2007}. These secondary electrons drive dissociation, ionisation, radical formation, and subsequent solid-state chemistry. The present irradiation conditions therefore provide an experimentally tractable analogue for probing electron-induced chemistry in EtA-containing ices. CASINO \citep{Casino_Drouin} simulations further indicate that 2~keV electrons deposit most of their energy within the overlying CH$_3$OH and H$_2$O layers while still allowing a fraction of the primary electrons to reach the upper region of the EtA layer, making this energy suitable for probing electron-induced processing in the present stratified geometry. During irradiation, additional mid-infrared spectra were acquired after 1, 5, 15, 30, and 60~min. This stepwise irradiation protocol was chosen to monitor the temporal evolution of the ice composition and the growth of irradiation-induced product bands. 

Upon completion of the irradiation, a temperature-programmed desorption (TPD) experiment was performed, during which mid-infrared spectra were collected at 10~K intervals between 20 and 310~K, with a ramp rate of 1~K/min. In addition, QMS signals were recorded during TPD to monitor desorbing species and fragments qualitatively. Subsequently, the irradiated and thermally processed residue remaining on the substrate was analysed \textit{ex situ} using electrospray ionisation mass spectrometry (ESI-MS). This analysis was carried out at the Chemical Instrumentation Laboratory -- LIQ, Faculty of Engineering, Design and Applied Sciences, ICESI University, Colombia.

\subsection{Spectral analysis and quantitative methodology}
\label{sec:spectralanalysis}

The column densities of the molecular components in the layered ice were calculated from the integrated absorbances of individual mid-infrared bands using standard infrared band strength analysis \citep{dHendecourt1986, Pilling2010, Muller2022}. The column density $N$ (molecules cm$^{-2}$) was obtained from:

\begin{equation}
N = \ln(10)\frac{S}{A}
\label{eq:column_density}
\end{equation}

\noindent where $S$ is the integrated absorbance (cm$^{-1}$) of a given band and $A$ is the corresponding band strength (cm molecule$^{-1}$). Once the column density of a deposited ice layer was determined, its apparent thickness $\theta$ ($\upmu$m) was calculated as:

\begin{equation}
\theta = 10{,}000 \frac{Nm}{\rho N_\mathrm{A}}
\label{eq:ice_thickness}
\end{equation}

\noindent where $m$ is the molar mass (g mol$^{-1}$), $\rho$ is the ice density (g cm$^{-3}$), and $N_\mathrm{A}$ is the Avogadro constant ($6.02\times10^{23}$ molecules mol$^{-1}$). The factor 10,000 is included to express $\theta$ in units of $\upmu$m.

For the layered samples, spectra recorded after the deposition of each individual component were used as references for spectral subtraction and comparison, allowing each deposited layer to be analysed separately. Band assignments were performed by comparison with literature spectra of EtA, H$_2$O, and CH$_3$OH (see Section~\ref{sec:results-layered-dep}).

Literature values for ice densities and band strengths were adopted for EtA, H$_2$O, and CH$_3$OH \citep{Reitmeier1940,suhasariaInfraredSpectraSolidstate2024,zhangSystematicIRVUV2024,gerakines2024sublimation,hudson2025infrared,hudsonInfraredBandStrengths2024}. The selected reference bands and the corresponding physical parameters used in the calculations are listed in Table~\ref{tab:PropertiesCalc}. These values were used to derive the column density of each deposited layer and, subsequently, to estimate its apparent thickness. Since the adopted band strengths and densities are taken from literature measurements and may depend on ice structure, temperature, and matrix environment, the resulting column densities and apparent layer thicknesses should be regarded as estimates. Following the uncertainty treatment used in related ICA studies \citep{Mifsud2023,Mifsud2024}, a 5~per cent uncertainty was assigned to the integrated absorbances. Because complete species-specific uncertainties were not available for all adopted band strengths and densities, a conservative overall relative uncertainty of 12.5~per cent was assigned to the derived column densities and apparent layer thicknesses.

\begin{table}
\centering
\caption{Physical parameters, integration regions, and infrared band strengths adopted for column density and apparent ice-thickness calculations.}
\label{tab:PropertiesCalc}
\footnotesize
\setlength{\tabcolsep}{8pt}
\renewcommand{\arraystretch}{1.1}
\begin{tabular}{@{}lllll@{}}
\hline
Species & $\rho$ & Int.\ region & Band & $A_{\rm ref}$ \\
 & (g\,cm$^{-3}$) & (cm$^{-1}$) & (cm$^{-1}$) & ($10^{-18}$\,cm\,molec.$^{-1}$) \\
\hline \hline
CH$_3$OH & 0.7791\textsuperscript{a} & 3650--2685 & 2958 & 152\textsuperscript{a} \\
H$_2$O   & 0.94\textsuperscript{b}   & 3100--3400 & 3271 & 200\textsuperscript{c} \\
EtA      & 1.01\textsuperscript{d}   & 1660--1560 &  1607    & 12$\pm$3\textsuperscript{e} \\
\hline
\end{tabular}

\vspace{0.4em}
\raggedright\footnotesize
\textbf{References:}
a)~\citet{hudsonInfraredBandStrengths2024};
b)~\citet{Bouilloud2015BIB};
c)~\citet{Gerakines1995};
d)~\citet{Reitmeier1940};
e)~\citet{suhasariaInfraredSpectraSolidstate2024}
\end{table}

\section{Results}
\label{sec:results}

\subsection{\texorpdfstring{Layered deposition of EtA--\ce{H2O}--\ce{CH3OH} ice analogue}{Layered deposition of EtA-H2O-CH3OH ice analogue}}
\label{sec:results-layered-dep}

\begin{figure}
\centering
\includegraphics[width=\columnwidth]{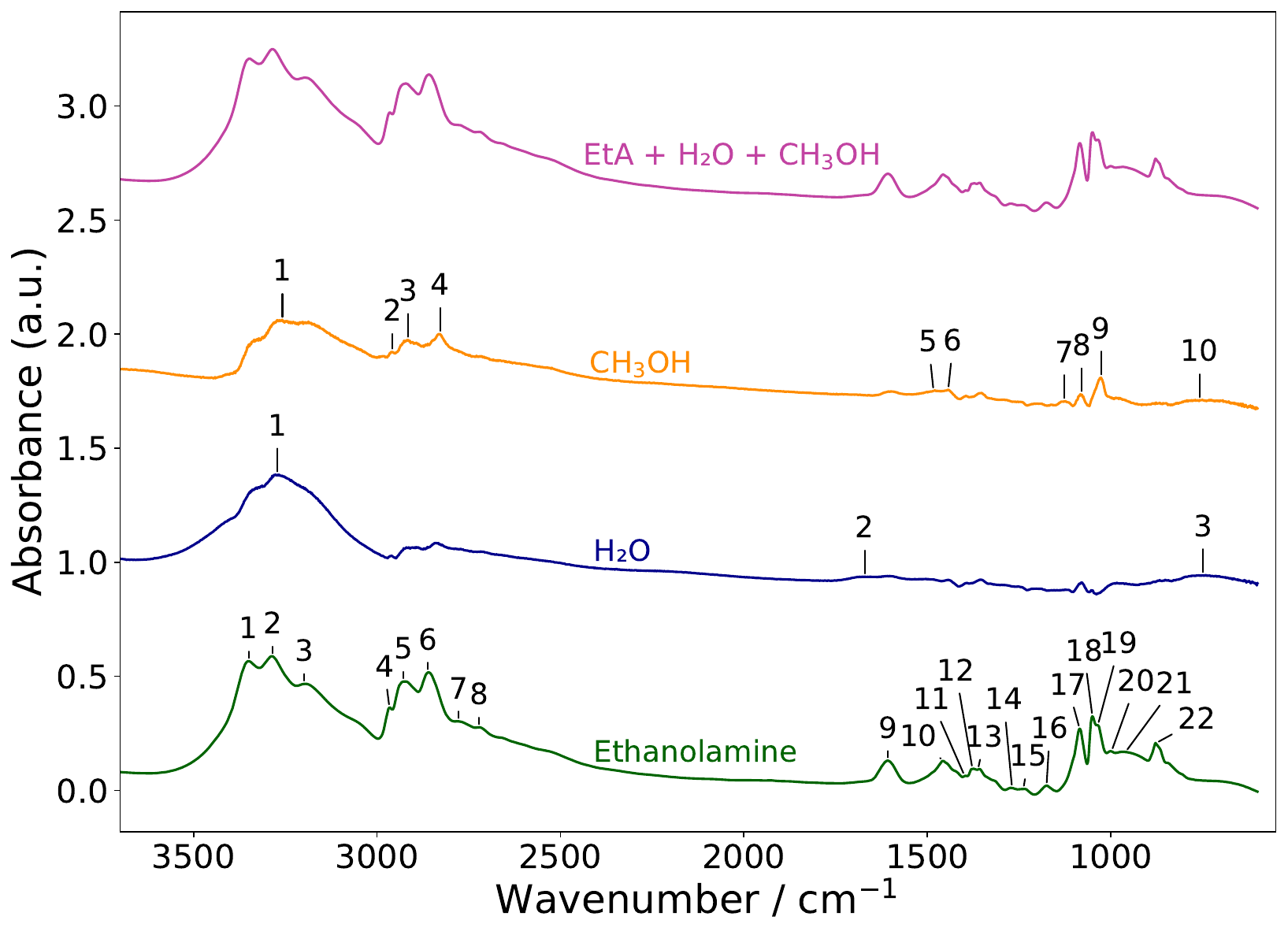}
\caption{Infrared spectra of the individual layers deposited at 20~K: ethanolamine (EtA; green), H$_2$O (blue), and methanol (CH$_3$OH; orange), together with the spectrum of the fully layered EtA--H$_2$O--CH$_3$OH ice (pink).}
\label{fig:EtA-Water-Methanol_layered}
\end{figure}

Figure~\ref{fig:EtA-Water-Methanol_layered} shows the mid-infrared spectra of the individual EtA, H$_2$O, and CH$_3$OH layers deposited sequentially at 20~K, together with the spectrum of the complete layered ice. The spectrum of the composite ice is well reproduced by the superposition of the individual component spectra, indicating that the layered deposition protocol preserves the spectral signatures of each constituent and that no detectable spectral evidence of reaction is found during deposition at 20~K.

The EtA absorption features observed in the layered ice are in good agreement with previously reported solid-state spectra \citep{zhangSystematicIRVUV2024,suhasariaInfraredSpectraSolidstate2024,ramachandranExperimentalComputationalStudy2024}. Several EtA bands previously reported but left unassigned, including features at 2777, 2722, and 1050~cm$^{-1}$, are clearly observed under the present experimental conditions. In addition, absorption features at 2966, 1269, and 998~cm$^{-1}$ are detected here but have not been reported previously for solid EtA. A complete overview of the observed band positions and vibrational assignments for EtA, H$_2$O, and CH$_3$OH is given in Table~\ref{tab:Layered_Spectra}.

\begin{table}
 \caption{Observed mid-infrared bands and vibrational assignments for EtA, H$_2$O, and CH$_3$OH in the layered ice following deposition at 20~K. Band numbers refer to those depicted in Figure~\ref{fig:EtA-Water-Methanol_layered}.}
 \label{tab:Layered_Spectra}
\begin{tabular}{lccc}
  \hline
  Species & Band & Wavenumber (cm$^{-1}$) & Mode \\
 \hline
 \hline
 EtA$^\textbf{a}$  & 1 & 3350 & $\nu$\textsubscript{as}(NH\textsubscript{2}) \\
   & 2 & 3284 & $\nu$\textsubscript{s}(NH\textsubscript{2}) \\
   & 3 & 3198 & $\nu$(O--H)\\
  &  4 & 2966 & -- \\
  &  5 & 2928 & $\nu$\textsubscript{as}(CH\textsubscript{2}) \\
   & 6 & 2861 & $\nu$\textsubscript{s}(CH\textsubscript{2}) \\
   & 7 & 2777 & -- \\
   & 8 & 2722 & -- \\ 
   & 9 & 1607 & $\delta$(NH\textsubscript{2}) \\
   & 10 & 1457 & $\delta$(CH\textsubscript{2}) \\
  & 11 & 1398 & $\nu$(C--C) $\delta$(OH, CH) \\
  &  12 & 1377 & $\omega$(CH\textsubscript{2}) \\
  &  13 & 1362 & $\tau$(NH, OH) \\
  &  14 & 1269 & -- \\
  &  15 & 1238 & $\tau$(CH\textsubscript{2}) \\
  &  16 & 1175 & $\rho$(CH\textsubscript{2}) \\
  &  17 & 1085 & $\nu$(C--O) \\
  &  18 & 1050 & -- \\
  &  19 & 1035 & $\nu$(C--N) \\
  &  20 & 998 & -- \\
  &  21 & 960 & $\omega$(NH\textsubscript{2}) \\
  &  22 & 877 & $\nu$(C--C) \\
 \hline
 H\textsubscript{2}O$^\textbf{b}$ &  1 & 3271 & $\nu$(O--H) \\
  &  2 & 1670 & $\delta$(H--O--H) \\
   & 3 & 748 & molecular librations \\
 \hline
 CH$_3$OH$^\textbf{c}$ &  1 & 3258 & $\nu$(O--H) \\
     & 2 & 2958 & $\nu$\textsubscript{as}(C--H) \\
     &3 & 2915 & -- \\
   & 4 & 2828 & $\nu$\textsubscript{s}(C--H) \\
   & 5 & 1483 & deformation mode \\
   & 6 & 1443 & deformation mode \\
   & 7 & 1127 & $\rho$(CH\textsubscript{3}) \\
   & 8 & 1079 & --   \\
   & 9 & 1026 & $\nu$(C--O) \\
   & 10 & 758 & $\tau$ \\
    \hline
    \hline
\end{tabular}
\vspace{0.2em}
\raggedright \footnotesize
\textbf{Modes:} $\nu$ (stretching), $\delta$ (bending), $\omega$ (umbrella), $\rho$ (rocking), $\tau$ (twisting).\\ 
$^\textbf{a}$Assignments based on the previous studies of \citet{suhasariaInfraredSpectraSolidstate2024}, \citet{ramachandranExperimentalComputationalStudy2024}, and \citet{zhangSystematicIRVUV2024}.\\
$^\textbf{b}$Assignments based on the previous study of \citet{hudson2025infrared}.\\
$^\textbf{c}$Assignments based on the previous studies of \citet{Ehrenfreund2000} and \citet{hudsonInfraredSpectroscopicPhysical2024}.
\end{table}

To quantify the relative abundances of the individual ice components, column densities were derived from the integrated band areas following the procedure described in Section~\ref{sec:spectralanalysis}. The calculated column densities are $N_{\mathrm{EtA}} = (3.6\pm0.5)\times10^{17}$~cm$^{-2}$, $N_{\mathrm{H_2O}} = (4.6\pm0.6)\times10^{16}$~cm$^{-2}$, and $N_{\mathrm{CH_3OH}} = (6.7\pm0.8)\times10^{16}$~cm$^{-2}$. The corresponding layer thicknesses were calculated using the equations described in Section~\ref{sec:spectralanalysis} and the physical parameters listed in Table~\ref{tab:PropertiesCalc}. The resulting apparent layer thicknesses are approximately $840\pm100$, $34\pm4$, and $104\pm13$~nm for EtA, H$_2$O, and CH$_3$OH, respectively. Thus, the H$_2$O and CH$_3$OH layers form a combined overlayer of approximately $140\pm20$~nm above the substantially thicker EtA deposit, giving a total apparent layered ice thickness of approximately $980\pm120$~nm.

\subsection{Irradiation studies}

\begin{figure}
\centering
\includegraphics[width=\columnwidth]{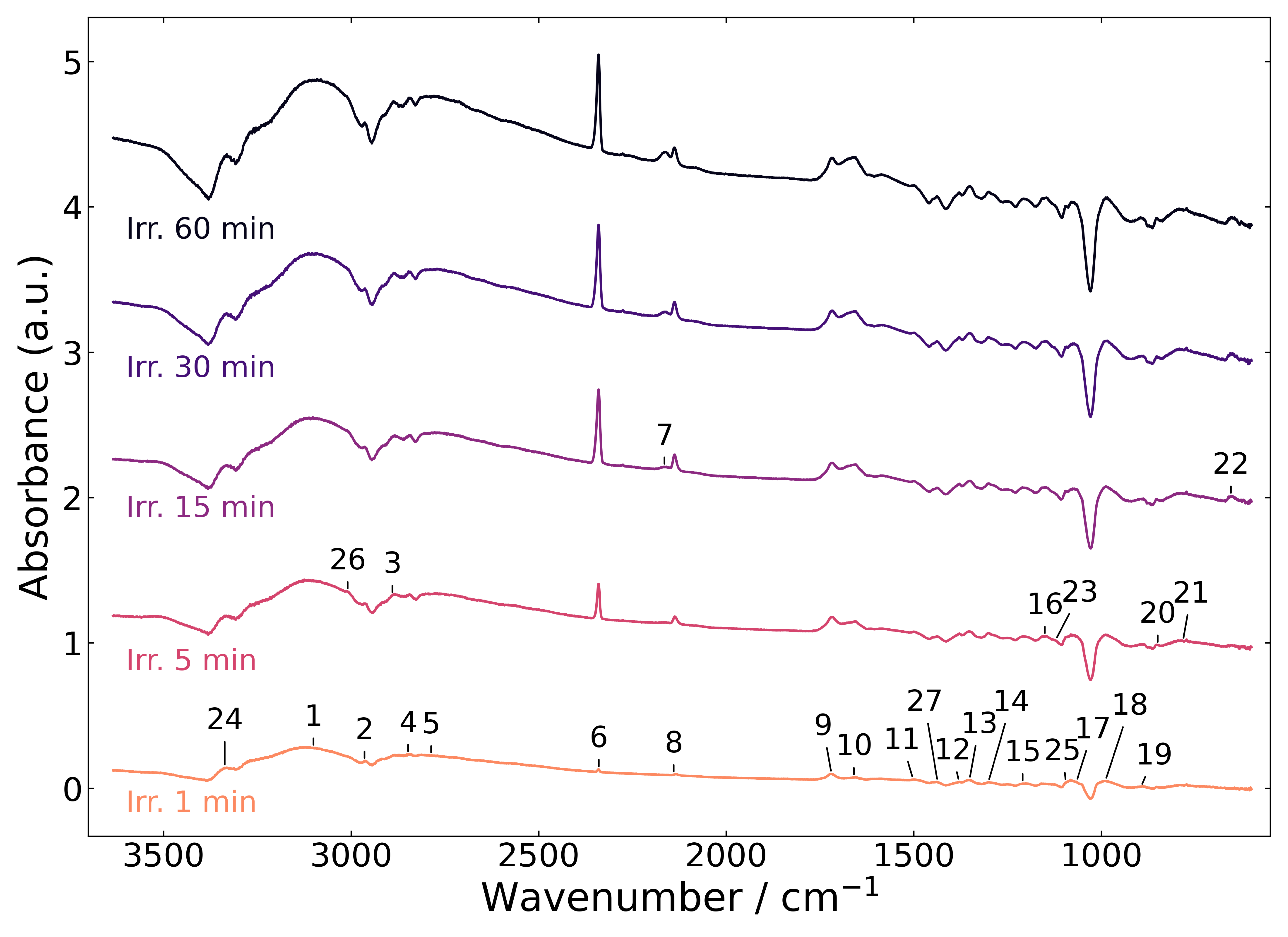}
\caption{Infrared spectra recorded during 2~keV electron irradiation of the EtA--H$_2$O--CH$_3$OH ice at 20 K, reaching a final fluence of $1.15 \times 10^{17}$ electrons cm$^{-2}$ over a duration of 60~min. Colours indicate increasing irradiation time from 1~min (yellow) to 60~min (black). Negative peaks correspond to the consumption/destruction of species and positive peaks to their formation.}
\label{fig:Ice-Irradiation}
\end{figure}

Electron irradiation of the layered EtA--H$_2$O--CH$_3$OH ice leads to the appearance and growth of numerous new infrared absorption features, indicating the formation of a diverse set of radiolysis products during irradiation (Figure~\ref{fig:Ice-Irradiation}). The spectra are shown as difference spectra relative to the unprocessed ice before irradiation; negative bands indicate depletion of parent or pre-existing species, whereas positive bands correspond to irradiation-induced products.

Owing to the complexity of the irradiated ice and extensive band overlap, not all observed features can be assigned unambiguously. This limitation is common in processed molecular ices, where irradiation can produce multiple species with similar functional groups and overlapping infrared bands \citep{schmidtElectronInducedProcessingMethanol2021}. Moreover, deviations between the band positions observed here and literature values are expected in molecular ice matrices. Previous studies of HCOOH-containing ices, H$_2$O$_2$ mixtures, and layered or mixed CH$_3$OH ice analogues have shown that peak positions and band profiles can depend strongly on ice structure, temperature, matrix composition, and hydrogen-bonding interactions \citep{bisschopInfraredSpectroscopyHCOOH2007,oberg2007,mullerSpectroscopicMeasurements2021,Oliveira2025}. The observed bands, possible assignments, contributing species, vibrational modes, and literature references are summarised in Table~\ref{tab:Product_table_Irr}. Formation pathways for the assigned and tentatively assigned species are discussed in Section~\ref{detection}.

The observed bands and their possible assignments are summarised in Table~\ref{tab:Product_table_Irr}. Many of the features are consistent with species previously reported in electron-irradiation studies of methanol ice \citep{Sullivan2016_MNRAS_CH3OH_electrons,mifsudElectronIrradiationThermal2021,schmidtElectronInducedProcessingMethanol2021} and ethanolamine-containing ices \citep{zhangSystematicIRVUV2024}, suggesting that common fragmentation and recombination pathways are active in the present layered ice system.

Prominent radiolysis products include CO$_2$, CO, H$_2$CO, CH$_4$, H$_2$O$_2$, and NH$_3$. The CO$_2$ asymmetric stretching mode at 2340~cm$^{-1}$ and the bending mode at 655~cm$^{-1}$ agree well with previously reported values for solid CO$_2$ \citep{ehrenfreundLaboratoryStudiesThermally,Isokoski2013_AA_CO2_ice_HR}. Similarly, the CO stretching band at 2140~cm$^{-1}$ matches literature assignments for radiolytically produced CO in molecular ices \citep{Pilling2010}. The feature at 1720~cm$^{-1}$ is consistent with the C=O stretching mode of H$_2$CO, although contributions from HCOOH and CH$_3$CHO cannot be excluded. Moreover, the CH$_4$ stretching and deformation bands at 3009~cm$^{-1}$ and 1302~cm$^{-1}$ are in good agreement with literature values \citep{hudsonActivationWeakIR2015,karteyevaInfraredSpectraMethanecontaining2026}. Vibrational features attributable to H$_2$O$_2$ are observed at 2848 and 1352~cm$^{-1}$ \citep{ZhengHydrogenPeroxide2006, IoppoloWaterFormationLow2010}, together with bands associated with NH$_3$ at 1067 and 3337~cm$^{-1}$ \citep{sandfordCondensationVaporizationStudies1993}.

\begin{table*}
 \caption{Infrared bands observed during 2~keV electron bombardment of the EtA--H$_2$O--CH$_3$OH ice and their possible assignments. Polyoxymethylene (POM), $N$,$N$-dimethylethanolamine (DMEtA), and N/O-bearing species (N/O-bear. spec.) are included as tentative contributors where relevant. Band numbers refer to those depicted in Figure~\ref{fig:Ice-Irradiation}.}
 \label{tab:Product_table_Irr}
\begin{tabular}{lccccc}
  \hline
  Band & Product & Wavenumber (cm$^{-1}$) & Mode & Contribution & References \\
  \hline
  \hline
  1  & H$_2$O   & 3100 & $\nu$(O--H)   &                      & a \\
  10 &          & 1660 & $\delta$(H--O--H) & $\nu_{\mathrm{HCOOH}}$(C=O) & a,b \\
  21 &          & 783  & molecular librations &                      & a,b \\
  \hline
  6  & CO$_2$  & 2340 & $\nu$(C=O)    &                      & a,c \\
  22 &         & 655  & $\delta$(C=O) &                      & a,c \\
  \hline
  8  & CO      & 2140 & $\nu$(C$\equiv$O)   &                      & b \\
  \hline
  9  & H$_2$CO & 1720 & $\nu$(C=O)    & $\nu_{\mathrm{HCOOH}}$(C=O), $\nu$(C=O)$_{\mathrm{CH_3CHO}}$  & d \\
  11 &         & 1502 & $\delta$(C--H) &                      & e \\
  \hline
  7  & OCN$^{-}$ & 2165 & $\nu$(C--N)  &                      & f \\
  14 &          & 1302 & $\nu$(C--O)  &  $\delta_{\mathrm{CH_4}}$             & f \\
  15 &          & 1210 & $\nu$(C--O)  & $\nu_{\mathrm{HCOOH}}$(C--O), & f \\
   & & & & $\nu_\mathrm{N/O-bear. spec.}$(CN) $\rho_\mathrm{N/O-bear. spec.}$(CH$_2$) & \\
  
  \hline
  4  & H$_2$O$_2$ & 2848 & combination &                      & g \\
  13 &           & 1352 & $\delta_{\mathrm{asym.}}$ & $\omega_{\mathrm{N/O-bear. spec.}}$(CH$_2$),  & g \\
  & & & & $\delta_{\mathrm{CH_3CHO}}$(CH$_3$)$-\omega$(C--H) & \\
  \hline
  2  & HCOOH & 2964 & $\nu$(C--H) & & h \\
  9  &       & 1720 & $\nu$(C=O)  & $\nu_{\mathrm{H_2CO}}$(C=O), $\nu$(C=O)$_{\mathrm{CH_3CHO}}$ & h \\
  10 &       & 1660 & $\nu$(C=O)  & $\delta_{\mathrm{H_2O}}$(H--O--H) & h \\
  12 &       & 1380 & $\nu$(O--H) & $\delta_{\mathrm{NH_2CHO}}$(C--H) & h \\
  15 &       & 1210 & $\nu$(C--O) & $\nu_{\mathrm{OCN^-}}$(C--O), & h \\
   & & & & $\nu_\mathrm{N/O-bear. spec.}$(CN) $\rho_\mathrm{N/O-bear. spec.}$(CH$_2$) & \\
  
  17 &       & 1067 & $\nu$(C--H) & $\omega_{\mathrm{NH_3}}$ & h \\
  \hline
   23 & CH$_3$OH & 1125 & $\rho$(CH\textsubscript{3})& & q\\
  \hline
  19 & N/O-bear. spec. & 895  & comb. $\delta$(N--(CH$_3$)$_3$) + $\delta$(N--(CH$_3$)$_3$) & & i \\
  16 &         & 1151 & $\rho$(CH$_3$) & & i,j \\
  20 &         & 850  & $\nu$(N--CH$_3$) & & i \\
  13 &         & 1352 & $\omega$(CH$_2$) & $\delta_{\mathrm{H_2O_2}}$\textsubscript{asym.}, $\delta_{\mathrm{CH_3CHO}}$(CH$_3$)$-\omega$(C--H) & i,x \\
  3 &          & 2890 & $\nu$(C--H) &  & t,u \\
  12 &         & 1380 & $\delta$(C--H) & $\nu_\mathrm{HCOOH}$(O--H) & t,u \\
  11 &         & 1502 & $\delta$(N--H$_3$) &                      & w \\
  15 &       & 1210 & $\nu$(CN) / $\rho$(CH$_2$) & $\nu_{\mathrm{OCN^-}}$(C--O), $\nu_{\mathrm{HCOOH}}$(C--O) & v\\
  \hline
  5  & POM & 2787 & $\nu$(C--H) & & d,k \\
  18 &     & 990  & $\nu$(C--O/C--N) & & l \\
  25 &     & 1095 & $\nu_{\mathrm{antisym.}}$(C--O--C)$-\delta$(O--C--O) & $\delta$(HCO) & k \\
  \hline
  25 & HCO & 1095 & $\delta$(HCO) & $\nu_{\mathrm{POM}}$\textsubscript{antisym.}(C--O--C)$-\delta_{\mathrm{POM}}$(O--C--O) & m \\
  \hline
  17 & NH$_3$ & 1067 & $\omega$ & $\nu$(C--H)$_{\mathrm{HCOOH}}$ & n \\
  24 &       & 3337 & $\nu$(N--H) & & n \\
  \hline
  27 & CH$_3$CHO & 1437 & deform(CH$_3$) & $\delta_{\mathrm{NH_4^+}}$ & o \\
  9  &          & 1720 & $\nu$(C=O) & $\nu_{\mathrm{HCHO}}$(C=O), $\nu_{\mathrm{HCOOH}}$(C=O) & o \\
  13 &          & 1352 & $\delta$(CH$_3$)$-\omega$(C--H) & $\omega_{\mathrm{N/O}}$(CH$_2$), $\delta_{\mathrm{H_2O_2}}$\textsubscript{asym.} & o \\
   \hline
  27 & NH$_4^+$CN$^-$ & 1437 & $\delta$(N--H) & CH$_3$CHO & p \\
  \hline
  26 & CH$_4$ & 3009 & $\nu$ & & r,s \\
  14 &        & 1302 & $\delta$ & $\nu_{\mathrm{OCN^{-}}}$(C--O) & r,s \\
  \hline
  
\end{tabular}

\vspace{0.4em}
\raggedright\footnotesize
\textbf{References:}
a)~\citet{ehrenfreundLaboratoryStudiesThermally};
b)~\citet{Pilling2010};
c)~\citet{Isokoski2013_AA_CO2_ice_HR};
d)~\citet{schutteExperimentalStudyOrganic1993};
e)~\citet{vinogradoffFormaldehydeMethylamineReactivity2013};
f)~\citet{mateCYANATEIONCOMPACT2012};
g)~\citet{ZhengHydrogenPeroxide2006,IoppoloWaterFormationLow2010};
h)~\citet{bisschopInfraredSpectroscopyHCOOH2007};
i)~\citet{pawlukojcINSDFTTemperature2014};
j)~\citet{StokrConformationDimethylaminoehtanol1987};
k)~\citet{LeRoyPOM2012};
l)~\citet{butscherRadicalinducedChemistryVUV2016};
m)~\citet{bennettLABORATORYSTUDIESFORMATION2011};
n)~\citet{Giuliano2014};
o)~\citet{TerwisschaVanScheltinga2018AandA};
p)~\citet{gerakines2024sublimation};
q)~\citet{hudsonInfraredSpectroscopicPhysical2024};
r)~\citet{hudsonActivationWeakIR2015};
s)~\citet{karteyevaInfraredSpectraMethanecontaining2026};
t)~\citet{brucatoInfraredStudyPure2006};
u)~\citet{slavicinskaHuntFormamideInterstellar2023};
v)~\citet{joshiChemicalLinkMethylamine2022};
w)~\citet{holtomCombinedExperimentalTheoretical2005};
x)~\citet{ciaravellaSynthesisComplexOrganic2019}.
\end{table*}

Bands corresponding to the cyanate anion (OCN$^{-}$) are detected at 2165, 1302, and 1210~cm$^{-1}$. These features are in good agreement with previous infrared assignments of OCN$^{-}$ in ice analogues \citep{mateCYANATEIONCOMPACT2012} and with observations of OCN$^{-}$ formation during electron irradiation of EtA and EtA--H$_2$O ices \citep{zhangSystematicIRVUV2024}. The detection of these bands suggests that N-bearing ionic species are efficiently produced under the present irradiation conditions. However, the features at 1302 and 1210~cm$^{-1}$ may include contributions from CH$_4$, HCOOH, and other N/O-bearing species.

\begin{figure}
\centering
\includegraphics[width=\columnwidth]{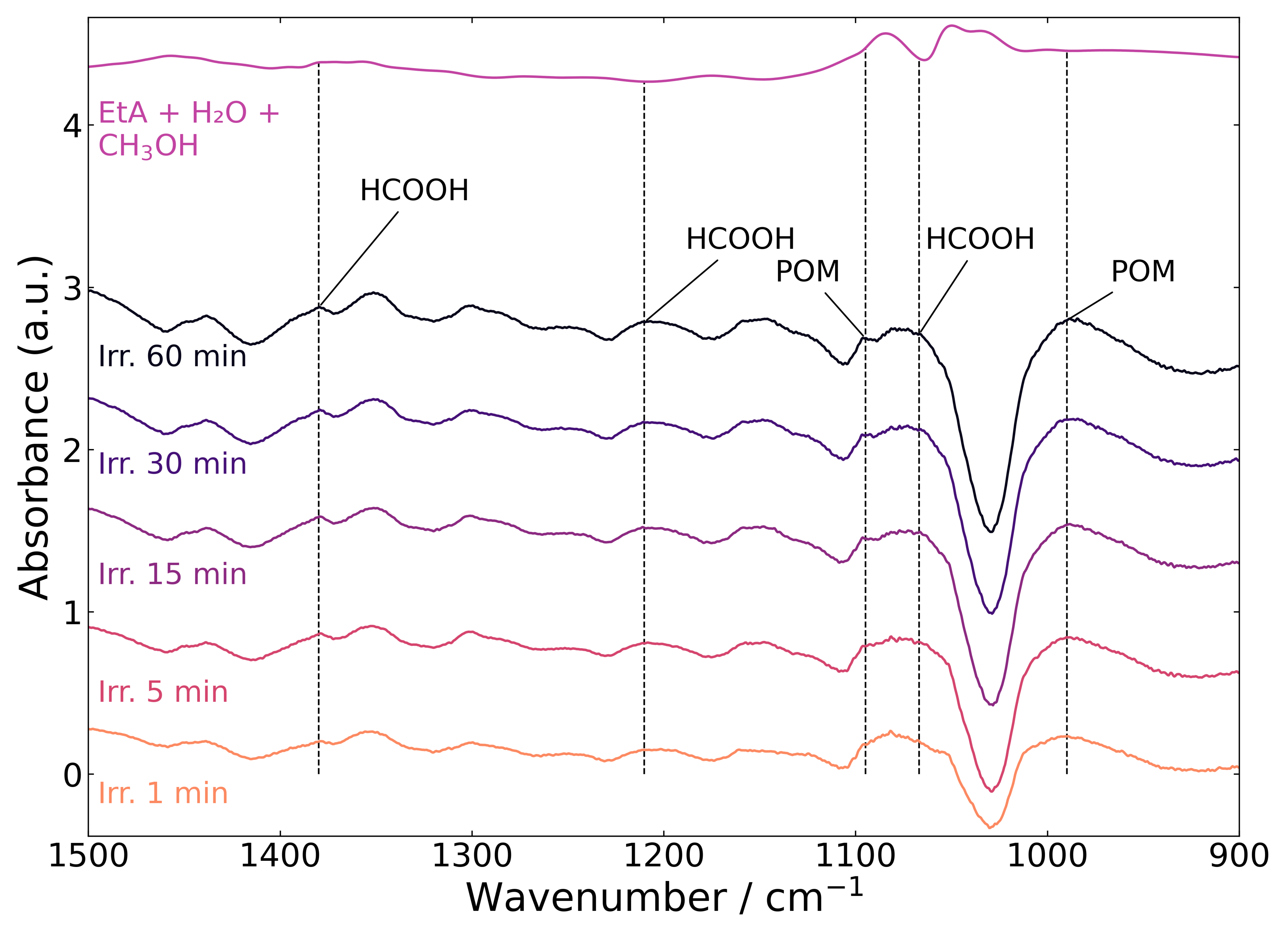}
\caption{Infrared difference spectra recorded during 2~keV electron irradiation of the EtA--H$_2$O--CH$_3$OH ice at 20~K. For clarity, all difference spectra are multiplied by a factor of 55; the unprocessed EtA--H$_2$O--CH$_3$OH spectrum shown in pink is not scaled in the same way. Colours indicate increasing irradiation time from 1~min (yellow) to 60~min (black). The black dotted lines mark selected bands consistent with formic acid (HCOOH) and polyoxymethylene-like material (POM). Negative bands indicate depletion of parent or pre-existing species, whereas positive bands correspond to irradiation-induced products.}
\label{fig:Zoom_in_POM_FA}
\end{figure}

In addition to simple radiolysis products, several spectral features suggest the formation of more complex organic species. Bands consistent with formic acid (HCOOH) are observed at 2964, 1720, 1660, 1380, and 1210~cm$^{-1}$ \citep{bisschopInfraredSpectroscopyHCOOH2007}. Although some of these bands overlap with modes of other species, their concurrent appearance and systematic growth during irradiation support a tentative HCOOH assignment (Figure~\ref{fig:Zoom_in_POM_FA}). In particular, the C--H stretching mode near 2964~cm$^{-1}$ (band number 2 in Figure~\ref{fig:Ice-Irradiation}) increases steadily with irradiation time, consistent with the HCOOH assignment \citep{bisschopInfraredSpectroscopyHCOOH2007}, although a contribution from formamide (NH$_2$CHO) in this spectral region cannot be excluded \citep{ciaravellaSynthesisComplexOrganic2019}. The strong growth observed near 1660~cm$^{-1}$ likely includes contributions from both HCOOH and the H$_2$O bending mode \citep{ehrenfreundLaboratoryStudiesThermally,bisschopInfraredSpectroscopyHCOOH2007,Pilling2010}. The band at 1380~cm$^{-1}$ may also receive contributions from NH$_2$CHO, as this region overlaps with modes reported for formamide-containing and amorphous formamide ices at low temperatures \citep{brucatoInfraredStudyPure2006,ciaravellaSynthesisComplexOrganic2019,slavicinskaHuntFormamideInterstellar2023}. A shoulder feature at 1677~cm$^{-1}$ is observed, but it was not included in the quantitative analysis.

A possible contribution from NH$_2$CHO is further suggested by the appearance of a band near 2890~cm$^{-1}$, close to the C--H stretching mode reported for formamide-containing ices \citep{ciaravellaSynthesisComplexOrganic2019,brucatoInfraredStudyPure2006,slavicinskaHuntFormamideInterstellar2023}. However, the assignment remains tentative because the corresponding in-plane C--H scissoring mode near 1380~cm$^{-1}$ overlaps with HCOOH and other products, while the asymmetric and symmetric N--H stretching modes of the NH$_2$ group are expected to be obscured by the broad H$_2$O stretching band. The band at 1210~cm$^{-1}$, although also compatible with the HCOOH assignment, may include a contribution from the aminomethyl radical ($^{\bm{\cdot}}$CH$_2$NH$_2$) \citep{joshiChemicalLinkMethylamine2022}.

Some features fall in spectral regions previously associated with glycine or its zwitterionic form. In particular, the band at 1502~cm$^{-1}$ lies close to the symmetric bending mode of the NH$_3^{+}$ group of zwitterionic glycine (H$_3$N$^{+}$CH$_2$COO$^{-}$) \citep{holtomCombinedExperimentalTheoretical2005}, and the feature at 1352~cm$^{-1}$ has also been reported in this context \citep{ciaravellaSynthesisComplexOrganic2019}. These bands are highly non-diagnostic in the present spectra, however, because they overlap with features attributable to H$_2$O$_2$, CH$_3$CHO, and choline-related species \citep{ZhengHydrogenPeroxide2006,IoppoloWaterFormationLow2010,pawlukojcINSDFTTemperature2014,TerwisschaVanScheltinga2018AandA}. Glycine is therefore considered only as a possible contributor and is not assigned unambiguously from the infrared data alone.

Other nitrogen-bearing features may be associated with methylated EtA derivatives and/or choline-like species. These include  a CH$_3$ rocking mode at 1151~cm$^{-1}$, an N--CH$_3$ stretching mode at 850~cm$^{-1}$, an umbrella deformation mode at 1352~cm$^{-1}$, and a combination band involving bending modes of N--(CH$_3$)$_3$ near 895~cm$^{-1}$. These features are compatible with reported choline spectra \citep{pawlukojcINSDFTTemperature2014}. In particular, the bands at 895 and 850~cm$^{-1}$ have been reported as characteristic of quaternary ammonium groups, with the former assigned to a combination of bending modes of N--(CH$_3$)$_3$ and the latter to the N--CH$_3$ stretching vibration \citep{pawlukojcINSDFTTemperature2014}. While these observations are suggestive, definitive identification of choline cannot be made from the infrared spectra alone because of spectral overlap and matrix effects.

Additional features are consistent with the formation of polyoxymethylene-like material (POM, HO--(CH$_2$--O)$_n$--H), as indicated by bands near 2787, 1095, and 990~cm$^{-1}$ \citep{schutteExperimentalStudyOrganic1993,LeRoyPOM2012,butscherRadicalinducedChemistryVUV2016}, as shown in Figure~\ref{fig:Zoom_in_POM_FA}. The feature near 1151~cm$^{-1}$ may also include a minor contribution from $N$,$N$-dimethylethanolamine (DMEtA) \citep{StokrConformationDimethylaminoehtanol1987}. 

\subsection{Temperature-programmed desorption}
\label{sec:SubsecTPD}

\begin{figure}
\centering
\includegraphics[width=0.9\columnwidth]{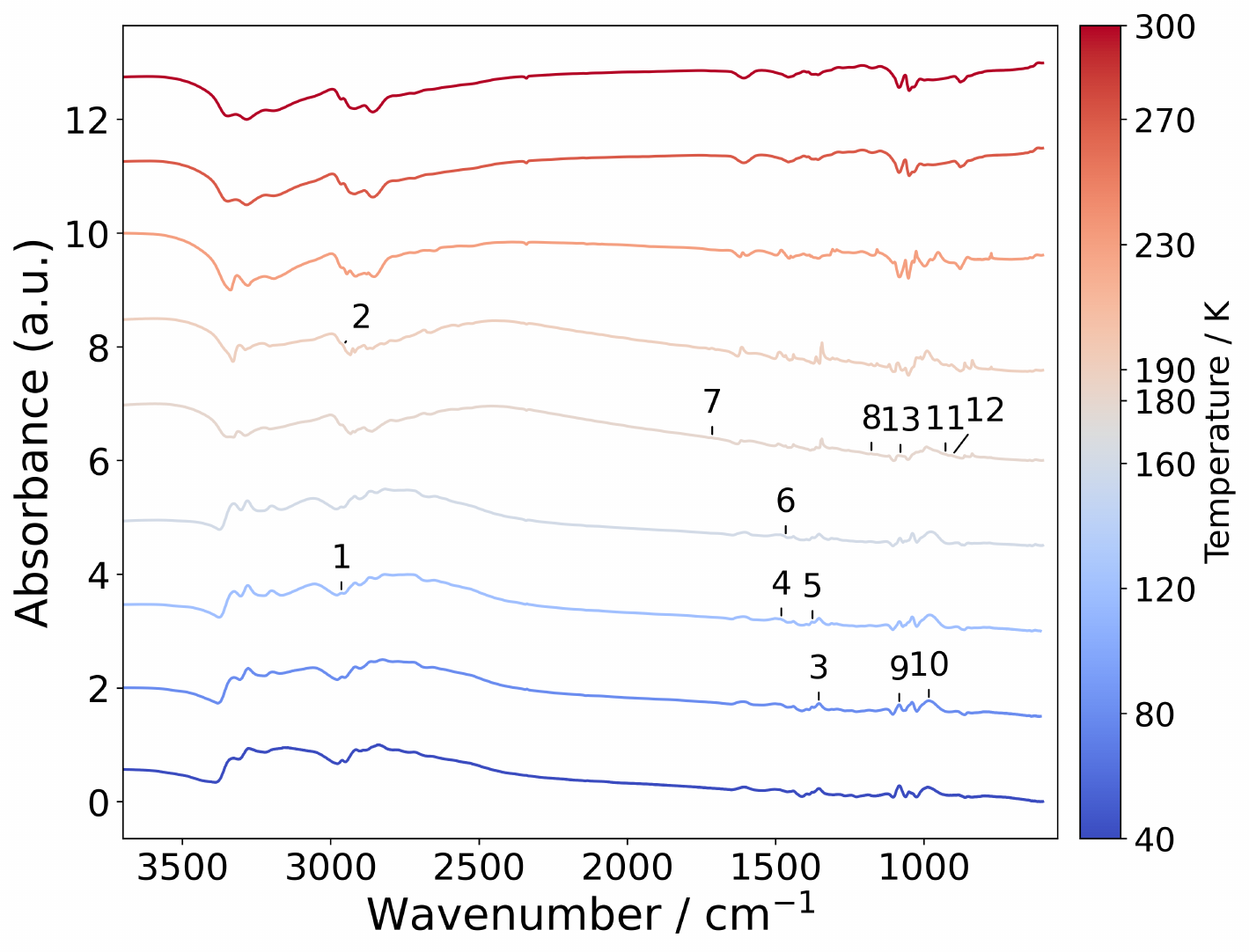}
\caption{Infrared spectra recorded during temperature-programmed desorption from 20~K to 310~K. For clarity, only nine of the thirty spectra are shown, and the spectra are displayed with an offset in absorbance. The assignment of each band is listed in Table~\ref{tab:Product_table_TPD}.}
\label{fig:Ice-TPD}
\end{figure}

Figure~\ref{fig:Ice-TPD} presents infrared spectra recorded during temperature-programmed desorption (TPD) of the irradiated EtA--H$_2$O--CH$_3$OH ice between 20 and 310~K. As the temperature increases, several absorption features associated with species formed during electron irradiation remain detectable and show temperature-dependent changes, consistent with the possible assignments listed in Table~\ref{tab:Product_table_TPD}. Rather than indicating the formation of an entirely new set of products during TPD, these spectra mainly trace the thermal response and structural reorganisation of the irradiated ice.

The tentative assignments of the N/O-bearing bands were guided by comparison with reference spectra of aqueous choline chloride solution. Choline chloride was used as a reference instead of pure choline, because choline is a cation and requires a counter-anion to be a stable compound. Chloride is well suited for this, as it is IR-silent in the mid-infrared and therefore does not interfere with the infrared spectrum of choline. Choline, [HOCH$_2$CH$_2$N(CH$_3$)$_3$]$^{+}$, retains the two-carbon hydroxyethyl backbone of ethanolamine, HOCH$_2$CH$_2$NH$_2$, but contains a fully N-methylated quaternary ammonium group. It can therefore be regarded as a structurally related reference for N/O-bearing products that could arise from progressive methylation of the ethanolamine amino group in the processed ice. These reference spectra are shown in Appendix Figure~\ref{fig:Choline} and Table~\ref{tab:Choline-Tab} and support the tentative association of bands near 2958, 1481, 1355, and 1084~cm$^{-1}$ with N/O-bearing species displaying choline-like spectral signatures.

\begin{table}
\caption{Infrared bands observed to persist during TPD of the irradiated EtA--H$_2$O--CH$_3$OH ice and their possible assignments. Band numbers refer to those labelled in Figure~\ref{fig:Ice-TPD}.}
\label{tab:Product_table_TPD}
\footnotesize
\setlength{\tabcolsep}{3pt}
\renewcommand{\arraystretch}{1.1}
\begin{tabular}{@{}l p{0.30\columnwidth} c p{0.34\columnwidth}@{}}
\hline
Band & Product & Wavenumber & Mode \\
     &         & (cm$^{-1}$) &      \\
\hline
\hline
1 & HCOOH$^\textbf{a}$ & 2964 & $\nu$(C--H) \\
\hline
2 & N/O-bear. spec.$^\textbf{b,g}$ & 2958 & C--H-related feature \\
3 &                              & 1355 & $\omega$(CH$_2$), $\nu_{\mathrm{s}}$(COO$^{-}$) \\
4 &                              & 1481 & $\delta$(CH$_3$) \\
9 &                              & 1084 & $\nu$(C--C), $\nu$(C--O) \\
\hline
5  & POM$^\textbf{c,d,f}$ & 1376 & -- \\
6  &                & 1466 & -- \\
10 &                & 984  & -- \\
11 &                & 928  & -- \\
12 &                & 906  & -- \\
13 &                & 1080 & -- \\
\hline
3 & HCOO$^{-}$$^\textbf{e}$ & 1355 & $\nu_{\mathrm{s}}$(COO$^{-}$) \\
\hline
7 & H$_2$CO$^\textbf{d,f}$ & 1714 & $\nu$(C=O) \\
8 &                 & 1178 & -- \\
\hline
\hline
\end{tabular}

\vspace{0.4em}
\raggedright\footnotesize
\textbf{References:}
$^\textbf{a}$\citet{bisschopInfraredSpectroscopyHCOOH2007};
$^\textbf{b}$\citet{pawlukojcINSDFTTemperature2014};
$^\textbf{c}$\citet{LeRoyPOM2012};
$^\textbf{d}$\citet{DuvernayForaldehydechemistr2014};
$^\textbf{e}$\citet{bennettLABORATORYSTUDIESFORMATION2011};
$^\textbf{f}$\citet{schutteExperimentalStudyOrganic1993};
$^\textbf{g}$Comparison with reference spectra of aqueous choline chloride solution; see Appendix Table~\ref{tab:Choline-Tab}.
\end{table}

Based on this comparison, bands tentatively associated with these N/O-bearing species become progressively sharper during warming up to approximately 80~K, particularly the features at 1352, 895, and 850~cm$^{-1}$, labelled as bands 13, 19, and 20 in Figure~\ref{fig:Ice-Irradiation}. This sharpening may reflect matrix reorganisation and increased local molecular ordering of the irradiated material. A contribution from phase separation or partial crystallisation cannot be excluded; however, the spectral overlap in this region prevents a definitive structural assignment.

In addition to these N/O-bearing features, a weak absorption band near 3195~cm$^{-1}$ emerges between 50 and 60~K. Although absorption in this region is often associated with O--H stretching modes of water, the observed band shape and temperature evolution differ from those expected for pure or mixed amorphous water ice \citep{oberg2007,mifsudLaboratoryExperimentsRadiation2022}. This suggests that the feature is unlikely to originate from bulk water and may instead arise from another OH-bearing species present in the processed material.

\begin{figure}
\centering
\includegraphics[width=0.95\columnwidth]{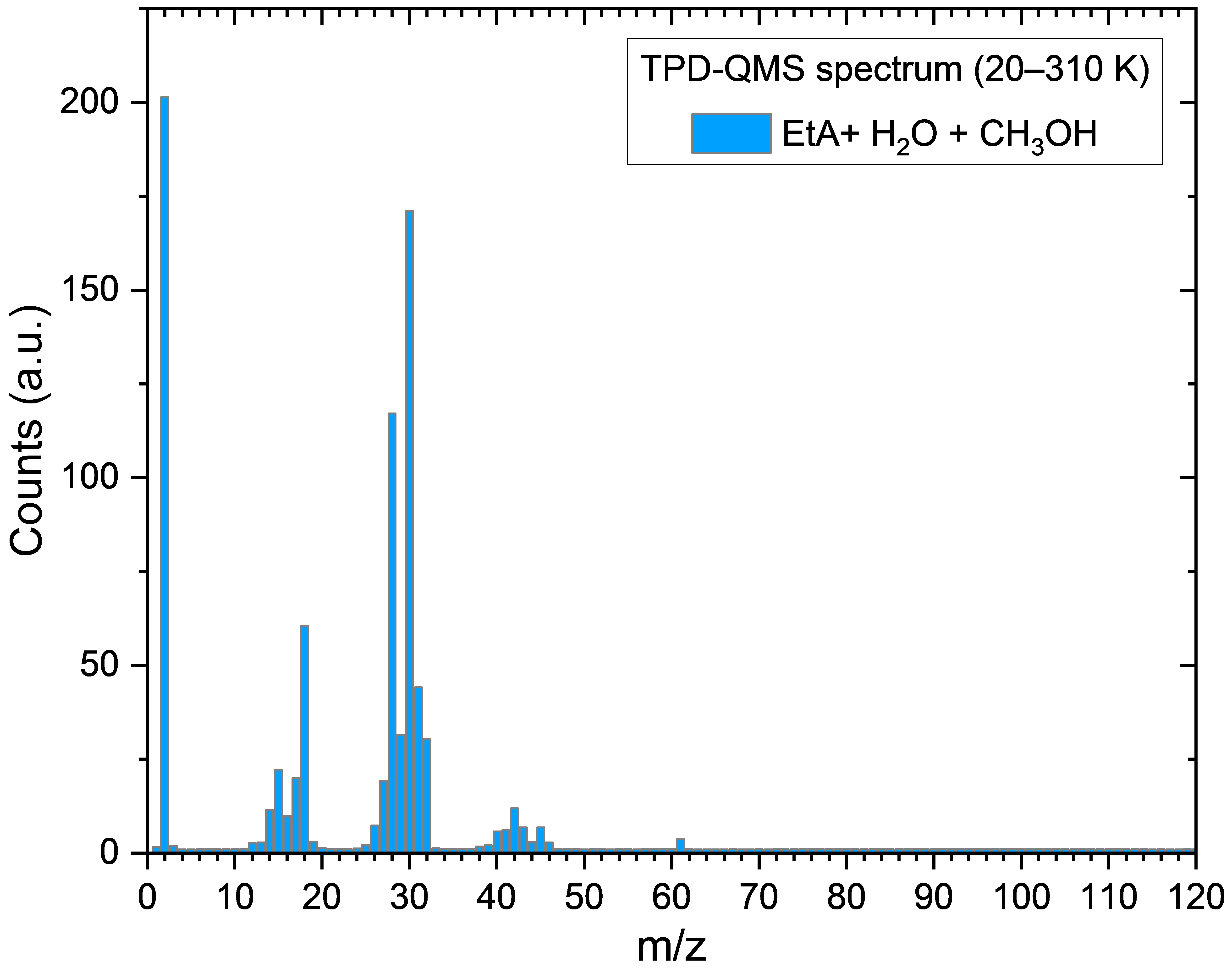}
\caption{Quadrupole mass spectrum recorded during temperature-programmed desorption of the irradiated EtA--H$_2$O--CH$_3$OH ice between 20 and 310~K.}
\label{fig:QMS-TPD}
\end{figure}

Figure~\ref{fig:QMS-TPD} shows the cumulative QMS ion counts recorded during TPD of the EtA--H$_2$O--CH$_3$OH ice from 20 to 310~K, following radiolytic processing. The QMS signal traces volatile species and ion fragments released into the gas phase during heating. The spectrum is dominated by low-mass ion signals, with prominent features below $m/z=50$ and only a weak higher-mass signal at $m/z=61$. The signal at $m/z=61$ is consistent with the nominal molecular ion of EtA and represents the largest $m/z$ value clearly detected in the spectrum. No significant signals are observed at higher $m/z$ values, indicating that volatile species heavier than EtA were not detected in the gas phase under the present experimental conditions. The lower-mass ion signals may arise from fragmentation of EtA during ionisation in the QMS and/or from volatile fragments associated with the processing of H$_2$O and CH$_3$OH. This behaviour is broadly consistent with previous time-of-flight mass spectrometry experiments on gas-phase EtA, which showed that the molecule readily fragments into lower-mass ions upon ionisation with synchrotron radiation \citep{quitian-laraPhotodissociationEthanolamineInterstellar2025}.

TPD analysis shows that several irradiation-induced features remain detectable in the solid residue over a broad temperature range and undergo clear spectral changes during warming. The IR spectra therefore trace the thermal evolution and reorganisation of this residue, whereas the QMS data indicate that the volatile fraction released during TPD is dominated by EtA and lower-mass fragments. These observations suggest that electron irradiation promotes both fragmentation and the formation of less volatile residue components whose infrared signatures become more distinct as the ice matrix reorganises during heating.

\subsection{\textit{Ex situ} ESI-MS analysis of the residue}
\label{sec:ESI}

\begin{figure}
\centering
\includegraphics[width=\columnwidth]{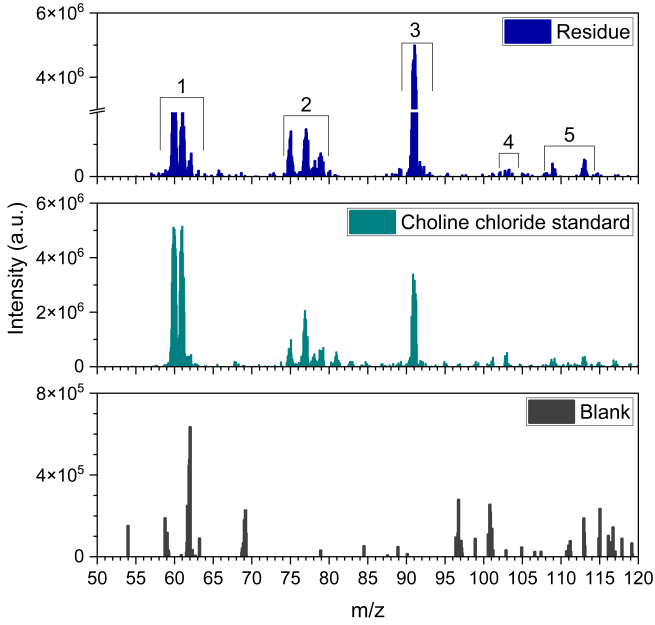}
\caption{ESI-MS spectra of the residue obtained after irradiation and TPD, choline chloride reference, and blank. For each sample, 286 scans were acquired under the same instrumental conditions; representative individual scans are shown. The residue spectrum is categorised into five families. Each family is numbered from 1-5, with increasing m/z.}
\label{fig:ESI-MS}
\end{figure}

\textit{Ex situ} analysis of the residue obtained after the irradiation and TPD experiments was carried out at the Chemical Instrumentation Laboratory – LIQ, 
Faculty of Engineering, Design and Applied Sciences, ICESI University, Colombia, using electrospray ionisation mass spectrometry (ESI-MS). Figure~\ref{fig:ESI-MS} shows the mass spectrum of the residue (top panel), a reference spectrum of choline chloride (middle panel), and the blank spectrum (bottom panel). For each sample, 286 scans were acquired under the same instrumental conditions. The spectra shown in Figure~\ref{fig:ESI-MS} correspond to representative individual scans selected from each dataset. The choline chloride standard was purchased from Sigma-Aldrich with a purity of 99 per cent, dissolved in ultrapure water, and introduced into the ESI source. The blank corresponds to a ZnSe substrate mounted in the same ICA sample holder during the experiment, but without direct EtA deposition or direct irradiation. This blank was kept in the same experimental environment and may therefore contain background deposition of H$_2$O and CH$_3$OH, as well as minor contributions from material deposited or redistributed during the experiment. The blank should therefore be regarded as a substrate/background control rather than as a full non-irradiated analogue of the layered EtA--H$_2$O--CH$_3$OH experiment. It allows possible contributions from the substrate, sample holder, background deposition, and sample-handling procedure to be assessed, but it cannot by itself isolate the effect of irradiation. The ESI-MS comparison with the blank is therefore used only as supporting evidence for the composition of the final residue, while the irradiation-induced chemistry is primarily established from the \textit{in situ} FTIR difference spectra recorded before and after electron bombardment of the same layered ice.

The residue spectrum shows a distribution of mass families that resembles the choline chloride reference spectrum more closely than the blank. In both the residue and the choline chloride reference, signals are grouped into five families across the analysed range, centred approximately at $m/z=60$--61, 75--80, 90--92, 103--106, and 112--116. This similarity is particularly evident in the relative distribution of the first three families, while the higher-$m/z$ families are weaker but still detectable in the residue. In contrast, the blank spectrum does not reproduce the same family pattern or relative intensity distribution. This comparison suggests that the processed residue contains N/O-bearing products with ESI-MS behaviour similar to that of methylated amino-alcohol or choline-like ionic species.

The first family, around $m/z=60$--61, is not diagnostic on its own, because several small nitrogen-bearing fragments may contribute in this region. In processed EtA ices, signals at $m/z=60$--61 have been associated with EtA-related ions, including the intact molecule at $m/z=61$ and an H-loss EtA fragment at $m/z=60$ \citep{Biancalani2024a}. However, this mass range is also characteristic of trimethylammonium-type fragments in choline-related compounds \citep{Holm2003,Wang2008}. Therefore, the presence of an intense family near $m/z=60$--61 in the residue is consistent with the formation of low-mass nitrogen-bearing and possibly methylated fragments, but it cannot be taken as direct evidence for choline.

The second and third families, near $m/z=75$--80 and $m/z=90$--92, indicate the presence of residue-derived ions at higher $m/z$ than the most prominent low-mass EtA-related fragments, although their assignments remain tentative. The family near $m/z=75$--80 is not unique to N/O-bearing species, since mass-spectrometric signals in this region have also been reported for oxygen-rich products formed during the energetic processing of methanol ice, including sugar-related species and their derivatives \citep{zhang2024}. Nevertheless, the overall distribution of the $m/z$ features, particularly the similarity between the five mass families of the residue and those of the choline chloride reference, suggests that the residue is not explained by methanol-derived C/H/O chemistry alone. The intense third family near $m/z=90$--92 remains unidentified, but a contribution from POM-related material cannot be excluded. During irradiation in the present experiments, H$_2$CO was detected, and recent studies on irradiated formaldehyde ice analogues have shown that energetic processing can lead to short-chain formaldehyde oligomers and POM-related products \citep{gong2025}. However, the prominence of this family in the residue and its similarity to the choline chloride reference point to a chemically complex residue; these characteristics are compatible with contributions from methylated amino-alcohol-type ionic species or choline-like species, but do not exclude other contributors in the same mass region. Given that the present ESI-MS data do not provide structural information, these related species can only be considered as possible contributors to the higher-$m/z$ components of the residue, rather than products identified with certainty.

The fourth family, around $m/z=103$--106, overlaps with the nominal mass region of the choline cation at $m/z=104$. However, because no isolated diagnostic signal is clearly resolved in the residue, this overlap is interpreted as part of a broader mass-family distribution resembling that of the choline chloride standard, rather than as a secure molecular identification. The highest-$m/z$ region is composed of two weak subfamilies centred approximately near $m/z=108$ and 114, which are also present in the choline chloride reference pattern but cannot be assigned to specific molecular carriers from nominal $m/z$ values alone. Their presence is nevertheless consistent with the formation of heavier products in the processed residue. This interpretation is supported by previous mass-spectrometric studies of processed EtA ices, where signals above the molecular mass of EtA, including masses above 120~u, were reported after irradiation \citep{Biancalani2024a}.
 
\section{Discussion}
\label{detection}

\begin{figure*}
\centering
\includegraphics[width=0.95\textwidth]{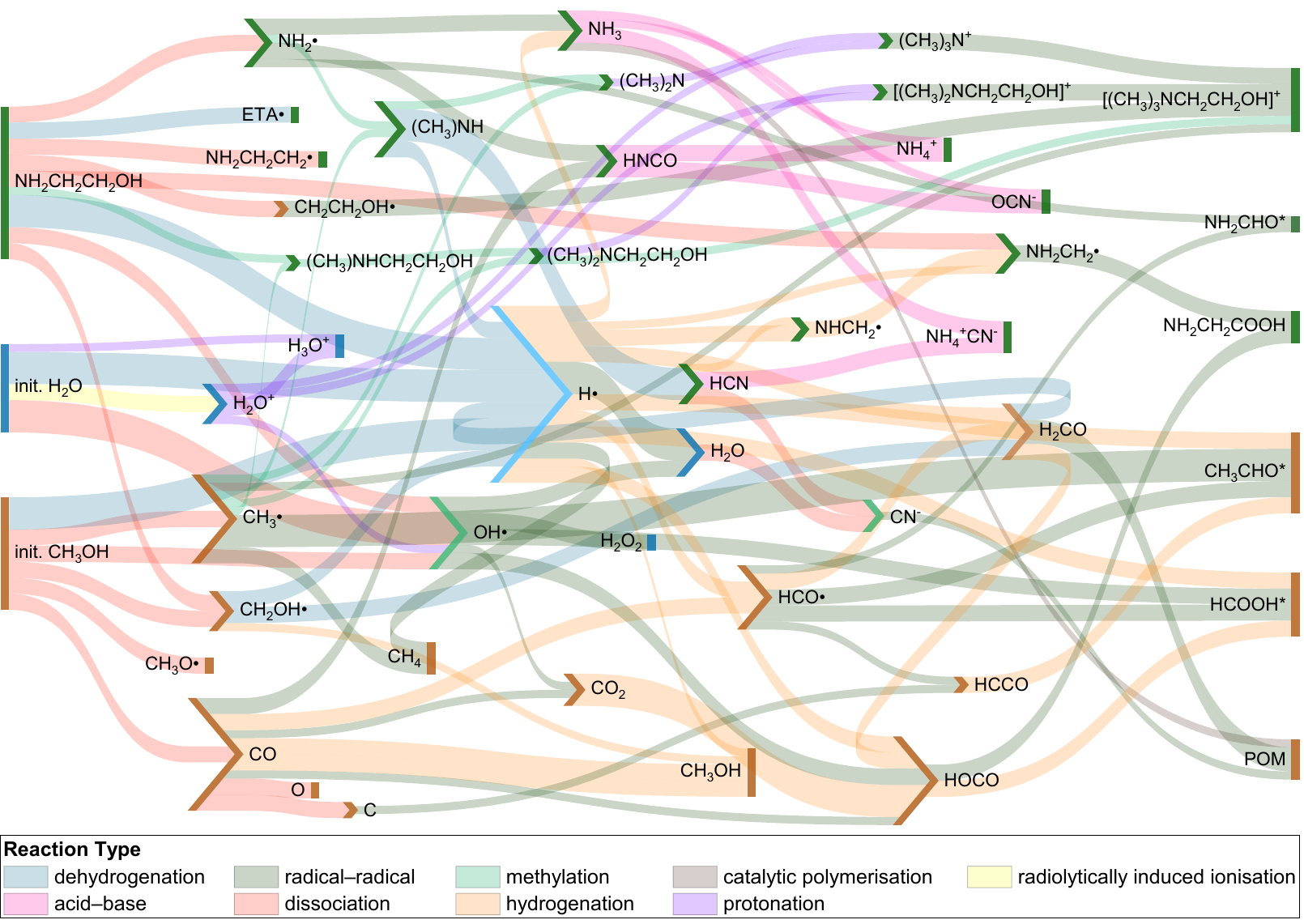}
\caption{Proposed qualitative reaction network for the EtA--H$_2$O--CH$_3$OH ice under 2~keV electron irradiation for 60~min. The scheme illustrates the main fragmentation pathways, radical and ionic intermediates, and products discussed in the text. The network is intended as a summary of plausible formation routes rather than a complete kinetic model. An asterisk (*) denotes tentatively assigned species; see text for details.}
\label{fig:Reaction_Scheme}
\end{figure*}

Electron irradiation of the layered EtA--H$_2$O--CH$_3$OH ice results in a chemically rich set of irradiation-induced features dominated by small O- and N-bearing species, together with tentative spectral evidence for less volatile and chemically more complex residue components. In this section, we discuss the assigned and tentatively assigned products in the context of established radiolysis pathways in astrophysical ice analogues, before addressing evidence for more complex reaction channels involving formic acid, POM-like material, and methylated amino-alcohol or choline-like mass-spectral features. A qualitative reaction network summarising the dominant pathways inferred from the present experiments is shown in Figure~\ref{fig:Reaction_Scheme}. 

The estimated apparent layer thicknesses are approximately 104~nm for CH$_3$OH, 34~nm for H$_2$O, and 837~nm for EtA. Since CH$_3$OH and H$_2$O were deposited above the EtA layer, the upper two layers have a combined thickness of approximately 138~nm. CASINO V2.48 simulations \citep{Casino_Drouin} indicate that 2~keV electrons have a maximum penetration depth of approximately 180~nm in the present ice configuration. Thus, around 90 per cent of the incident electrons are expected to be implanted within the CH$_3$OH and H$_2$O layers, while the remaining fraction can reach and interact with the upper region of the EtA layer. This depth distribution is consistent with the predominance of products commonly observed in processed methanol- and water-rich ices, while still allowing EtA-derived chemistry through the fraction of primary electrons reaching the upper EtA layer and through radical-mediated reactions at or near the H$_2$O/EtA interface. In addition, low-energy secondary electrons generated during the deceleration of the incident 2~keV electrons may contribute to further processing of the surrounding ice, as commonly discussed for irradiated molecular ices \citep{arumainayagamExtraterrestrialPrebioticMolecules2019}.

The pathways shown in Figure~\ref{fig:Reaction_Scheme} are represented by nodes and links. The colour of each node indicates the parent species or chemical family to which it is mainly related, while the colour of each link indicates the reaction type. The reactions are grouped into hydrogenation (light orange), dehydrogenation (bright blue), polymerisation (grey), acid--base reactions (light pink), radical--radical reactions (green), dissociation (red), methylation or methyl-transfer-type pathways (bright green), protonation (purple), and radiolytically induced ionisation (blue). H$^{\bm{\cdot}}$ and OH$^{\bm{\cdot}}$ are produced through several pathways and participate in multiple subsequent reactions; therefore, they are highlighted as central radical intermediates in the network.

\subsection{Formation of small and intermediate radiolysis products}

Most of the species assigned or tentatively assigned during irradiation, including CO$_2$, CO, H$_2$O$_2$, NH$_3$, OCN$^{-}$, H$_2$CO, NH$_4^+$CN$^-$, CH$_3$CHO, and HCO, are consistent with products commonly observed in electron-, proton-, and soft-X-ray-processed interstellar ice analogues dominated by water and methanol \citep{Sullivan2016_MNRAS_CH3OH_electrons,Pilling2019,mifsudElectronIrradiationThermal2021,mifsudProtonElectronIrradiations2023}. Their appearance in the present ternary system indicates that the inclusion of EtA does not suppress the primary radiolytic chemistry driven by H$_2$O and CH$_3$OH, but also allows additional nitrogen-bearing reaction channels. This interpretation is consistent with the QMS data recorded during TPD, which are dominated by low-mass volatile species and ion fragments.

Electron-induced fragmentation of EtA can proceed through cleavage of C--C, C--N, or C--O bonds, producing radicals such as CH$_2$OH$^{\bm{\cdot}}$, CH$_2$NH$_2^{\bm{\cdot}}$, NH$_2^{\bm{\cdot}}$, CH$_2$CH$_2$OH$^{\bm{\cdot}}$, and CH$_2$CH$_2$NH$_2^{\bm{\cdot}}$ \citep{sladkovaDestructionAminoAlcohols2014,zhangSystematicIRVUV2024,suhasariaInfraredSpectraSolidstate2024}. In Figure~\ref{fig:Reaction_Scheme}, these bond-cleavage processes are represented by red links and classified as dissociation pathways. More complex fragmentation processes involving the simultaneous cleavage of multiple bonds have also been reported, yielding fragments such as OH$^{\bm{\cdot}}$, NH$_2^{\bm{\cdot}}$, and C$_2$H$_4$ \citep{zhangSystematicIRVUV2024}.

Fragmentation of CH$_3$OH under electron irradiation likely produces primary radicals such as CH$_3^{\bm{\cdot}}$, CH$_2$OH$^{\bm{\cdot}}$, and CH$_3$O$^{\bm{\cdot}}$ \citep{schmidtElectronInducedProcessingMethanol2021,Herczku2021}. The formation of CO$_2$ has been discussed by \citet{Schmidt2019}, \citet{Herczku2021}, and \citet{mifsudProtonElectronIrradiations2023}, either through reactions between CO and atomic oxygen to form CO$_2$, or through OH-mediated oxidation of CO via HO--CO intermediates \citep{ioppoloSurfaceFormationCO22011,mifsudProtonElectronIrradiations2023}. In addition, CH$_3$OH may also be regenerated during irradiation through sequential hydrogenation of CO or through radical recombination between CH$_2$OH$^{\bm{\cdot}}$ and H$^{\bm{\cdot}}$ \citep{zhangSystematicIRVUV2024}.

Hydrogen peroxide (H$_2$O$_2$) is readily explained by radical--radical recombination of OH$^{\bm{\cdot}}$ fragments generated during water and EtA radiolysis \citep{mifsudLaboratoryExperimentsRadiation2022}, as indicated by the green links in Figure~\ref{fig:Reaction_Scheme}. Water itself can be regenerated during irradiation through recombination of H$^{\bm{\cdot}}$ atoms, released by dehydrogenation of EtA and CH$_3$OH, with OH$^{\bm{\cdot}}$ radicals. Radiolytically induced ionisation of H$_2$O during electron bombardment can also produce H$_2$O$^{+}$, followed by proton transfer to a neighbouring H$_2$O molecule to form H$_3$O$^{+}$ and an OH$^{\bm{\cdot}}$ radical \citep{arumainayagamExtraterrestrialPrebioticMolecules2019}, as indicated by the purple and blue links in Figure~\ref{fig:Reaction_Scheme}. 

Formaldehyde (H$_2$CO) bands are observed within the first minute of irradiation, consistent with rapid methanol dehydrogenation \citep{schmidtElectronInducedProcessingMethanol2021}, as indicated by the bright blue link in Figure~\ref{fig:Reaction_Scheme}. An additional contribution to H$_2$CO formation may arise from further processing of CH$_2$OH$^{\bm{\cdot}}$ fragments originating from EtA or CH$_3$OH dissociation \citep{Pilling2010, Butscher2017, Schmidt2019, zhangSystematicIRVUV2024}, as shown in Figure~\ref{fig:Reaction_Scheme}.

Acetaldehyde (CH$_3$CHO) may form either through radical--radical recombination between CH$_3^{\bm{\cdot}}$ and HCO$^{\bm{\cdot}}$, or through the reaction of HCO with atomic carbon to form HCCO, followed by sequential hydrogenation to yield CH$_3$CHO \citep{TerwisschaVanScheltinga2018AandA}. Since the infrared bands assigned to CH$_3$CHO overlap with other carbonyl-containing species, this pathway is included as a plausible contributor rather than as a unique assignment.

Finally, nitrogen-bearing species such as NH$_3$ and NH$_4^{+}$ are formed primarily through hydrogenation and proton-transfer reactions involving NH$_2^{\bm{\cdot}}$ radicals \citep{sladkovaDestructionAminoAlcohols2014}. HNCO may form through reactions involving CO and NH/NH$_2$-bearing fragments \citep{fedoseev2015,fedoseev2016}. The detection of OCN$^{-}$ is consistent with the well-established acid--base reaction between ammonia and isocyanic acid following energetic or thermal processing \citep{vanBroekhuizen2004,raunier2004}, as indicated by the pink links in Figure~\ref{fig:Reaction_Scheme}.

The presence of these nitrogen-bearing products indicates that EtA contributes to the observed chemistry, even though most of the deposited energy is expected to be absorbed within the overlying CH$_3$OH and H$_2$O layers. As discussed in Section~\ref{sec:SubsecTPD}, the tentative assignment of selected N/O-bearing infrared bands was guided by comparison with reference spectra of aqueous choline chloride solution. The corresponding reference spectra and band assignments are presented in Appendix Figure~\ref{fig:Choline} and Table~\ref{tab:Choline-Tab}. EtA contribution may arise from the fraction of primary electrons that reaches the upper EtA layer, as well as from low-energy secondary electrons and radicals generated within the irradiated ice \citep{arumainayagamExtraterrestrialPrebioticMolecules2019}. The nitrogen-bearing products observed during irradiation are therefore consistent with a limited but chemically relevant contribution from EtA processing, in agreement with the \textit{ex situ} ESI-MS analysis of the final residue discussed in Section~\ref{sec:ESI}.

\subsection{Formic acid formation}

Several infrared features observed at 2964, 1720, 1660, 1380, 1210, and 1067~cm$^{-1}$ are consistent with vibrational modes of formic acid (HCOOH). Although partial overlap with water and formaldehyde bands cannot be excluded, the correlated appearance of these features supports a tentative assignment to HCOOH.

Formic acid formation under energetic processing has been investigated in CO:H$_2$O and H$_2$O:CO$_2$ ice systems \citep{schutteWeakIceAbsorption1999,Pilling2010,bennettLABORATORYSTUDIESFORMATION2011}. In CO:H$_2$O ices, HCOOH formation proceeds through a mechanism involving (i) water dissociation into H$^{\bm{\cdot}}$ and OH$^{\bm{\cdot}}$, (ii) formation of the formyl radical via H$^{\bm{\cdot}}$ + CO, and (iii) barrierless recombination of HCO$^{\bm{\cdot}}$ with OH$^{\bm{\cdot}}$, consistent with the pathways summarised in Figure~\ref{fig:Reaction_Scheme}. All necessary precursors are present in the present EtA--H$_2$O--CH$_3$OH system, making this pathway viable without invoking EtA-specific chemistry. The tentative identification of HCOOH therefore highlights the role of water- and CO-derived radicals in the product inventory. Formic acid may also form through recombination of OH$^{\bm{\cdot}}$ and CO, or the addition of a hydrogen atom to CO$_2$ to form HOCO, followed by hydrogenation \citep{Goumans2008,bennettLABORATORYSTUDIESFORMATION2011,Schmidt2019}.

\subsection{Polyoxymethylene formation and thermal evolution}

Polyoxymethylene-like material formation is inferred from characteristic C--O and C--H vibrational features that emerge following irradiation and persist during subsequent heating. POM formation has been reported previously under UV photolysis and thermal processing of formaldehyde-containing ices \citep{schutteExperimentalStudyOrganic1993,bernsteinOrganicCompoundsProduced1995,butscherRadicalAssistedPolymerization2019}. In these experiments, ammonia has been proposed as a polymerisation catalyst \citep{schutteExperimentalStudyOrganic1993}. In the present experiment, the early appearance of H$_2$CO provides a natural precursor for polymerisation.

During temperature-programmed desorption, bands assigned to POM-like material remain detectable up to approximately 190~K, above which they disappear. This comparatively low thermal stability suggests the formation of short-chain POM oligomers, consistent with formation in a chemically complex ice matrix \citep{LeRoyPOM2012,DuvernayForaldehydechemistr2014}. Delayed sublimation of formaldehyde relative to its pure-ice desorption temperature (135~K; \citealt{nobleDesorptionH2COInterstellar2012}) likely reflects trapping within the mixed ice structure \citep{DuvernayForaldehydechemistr2014}. Thus, the POM-related features are best interpreted as evidence for formaldehyde-driven oligomerisation within the processed residue, rather than as a definitive assignment of a specific polymer chain length or distribution.

\subsection{N/O-bearing complex residue components and methylation pathways}

Several spectral features observed in the irradiated EtA--H$_2$O--CH$_3$OH ice, notably those near 895 and 850~cm$^{-1}$, are compatible with vibrations associated with quaternary ammonium or N-methylated species \citep{pawlukojcINSDFTTemperature2014}. These bands show similarities with aqueous choline chloride solution reference spectra presented in Appendix~Figure~\ref{fig:Choline} and Table~\ref{tab:Choline-Tab}. Their emergence during irradiation suggests that methylation of EtA-derived fragments could occur in the presence of CH$_3$OH under electron bombardment, although these assignments remain tentative.

Methanol radiolysis is known to generate CH$_3^{\bm{\cdot}}$ radicals \citep{schmidtElectronInducedProcessingMethanol2021}. Although these radicals are not directly observed in the present experiments because of their short lifetimes, their participation in secondary ice chemistry is well established. Sequential methylation of EtA-derived intermediates to $N$-methylethanolamine and $N$,$N$-dimethylethanolamine represents a chemically plausible pathway, analogous to methylation reactions reported in other irradiated amine-containing systems \citep{Pham2023_methyl_aniline}. In the present layered ice, such reactions would require spatial overlap between methanol-derived methyl radicals and EtA-derived amino-alcohol fragments, most likely near the CH$_3$OH/H$_2$O/EtA interfacial region or through secondary-electron-driven chemistry.

The final methylation step leading to a quaternary ammonium species such as the choline cation, however, is expected to be challenging in the solid phase. Formation of a stable quaternary ammonium cation likely requires either highly localised ion--molecule encounters or transient cationic intermediates. In this context, oxidation of dimethylethanolamine to a radical cation, potentially mediated by radiolytically induced ionisation of water to H$_2$O$^{+\bm{\cdot}}$, provides a conceivable alternative pathway \citep{arumainayagamExtraterrestrialPrebioticMolecules2019,ZhangAmbientCatalyst-free2024}. Reaction of such a radical cation with a nearby CH$_3^{\bm{\cdot}}$ radical could, in principle, provide a route towards the choline cation or a related quaternary-ammonium-type ion, as represented in Figure~\ref{fig:Reaction_Scheme}. This pathway should therefore be regarded as a proposed route rather than a confirmed mechanism.

The comparatively weak intensity of the choline-related infrared bands suggests that this channel is inefficient, consistent with the known radiosensitivity of choline reported in earlier irradiation studies \citep{Lindblom1961,Symons1971}. In addition, the limited penetration of 2~keV electrons means that only the upper region of the EtA layer is expected to be directly processed, while most of the deposited energy is absorbed within the overlying CH$_3$OH and H$_2$O layers. This restricted access to EtA may further limit the formation of methylated EtA-derived products, and rapid destruction of newly formed choline or choline-like species under continued irradiation may further suppress their net abundance. This interpretation is consistent with the ESI-MS data, where the fourth and fifth mass families show similarities with the choline chloride reference spectrum, as shown in Figure~\ref{fig:ESI-MS}. Although no isolated diagnostic signal at $m/z=104$ is clearly detected, the mass-family distribution of the residue resembles that of the choline chloride reference. This suggests that methylated amino-alcohol-type ions, choline-like ions, or related fragments may contribute to the residue, while a secure identification of choline itself cannot be made from the present data.  

Other N/O-bearing products cannot be excluded in the processed residue. For example, formamide (NH$_2$CHO) formation has been investigated by \citet{jonesMECHANISTICALSTUDIESPRODUCTION2011}, who proposed a pathway involving initial hydrogenation of CO to form the formyl radical, HCO$^{\bm{\cdot}}$, followed by recombination with the amino radical, NH$_2^{\bm{\cdot}}$. This route is relevant to the present experiments because both CO and NH$_2$-bearing fragments can be produced during irradiation of the EtA--H$_2$O--CH$_3$OH ice. Furthermore, glycine has been investigated in several experimental and theoretical studies under energetic processing conditions \citep{holtomCombinedExperimentalTheoretical2005,mateInfraredStudySolid2011,portugalRadiolysisAminoAcids2014,joshiChemicalLinkMethylamine2022}. However, in the present work, formamide- and glycine-related assignments should be regarded as tentative because of extensive band overlap in the infrared spectra and the lack of structural confirmation from the ESI-MS data.

\section{Astrochemical and Astrobiological Implications}

These experiments show that EtA, deposited beneath H$_2$O and CH$_3$OH layers, remains spectroscopically detectable after 60~min of 2~keV electron irradiation while participating in a network of radiation-driven reactions relevant to dense molecular cloud environments. The predominance of small molecules among the radiolysis products is consistent with efficient fragmentation and radical chemistry in the processed ice, whereas the persistence of EtA spectral signatures after irradiation suggests that, when embedded within stratified H$_2$O- and CH$_3$OH-rich grain mantles, a fraction of EtA may survive energetic processing and remain available for further chemical evolution during ice warming, desorption, or incorporation into larger icy bodies.

EtA has been detected in the GC molecular cloud G+0.693$-$0.027 \citep{rivillaDiscoverySpaceEthanolamine2021}, a chemically rich region where low-velocity shocks and enhanced cosmic-ray or X-ray processing have been proposed to play an important role in the observed molecular inventory \citep{zeng2018,zeng2020}. The present results show that, in a layered ice containing EtA, H$_2$O, and CH$_3$OH, EtA can participate in irradiation-driven chemistry leading to N/O-bearing products of increasing complexity. Although the present data do not provide secure evidence for efficient choline formation, and although choline-containing salts are known to be radiosensitive \citep{Lindblom1961,Symons1971}, partial methylation of EtA-derived fragments by CH$_3$OH-derived radicals remains a plausible route for expanding the chemical inventory available in processed interstellar ices \citep{schmidtElectronInducedProcessingMethanol2021, Pham2023_methyl_aniline}.

If preserved during subsequent thermal evolution and later incorporated into planetesimals or cometary material, such processed N/O-bearing residues could contribute to the molecular complexity delivered to young planetary systems. This possibility is consistent with studies linking interstellar organic chemistry, cometary and meteoritic material, and the exogenous delivery of prebiotic compounds to the early Earth \citep{ChybaSagan1992,Ehrenfreund2000,MummaCharnley2011}. The detection of prebiotic molecules in cometary material, including glycine, further supports the relevance of small-body reservoirs as carriers of molecular complexity \citep{Elsila2009, Altwegg2016}. While the present work does not demonstrate efficient synthesis of fully formed biomolecules, it shows that EtA-containing ices are chemically active systems capable of supporting complex nitrogen- and oxygen-bearing chemistry under energetic processing, consistent with previous laboratory studies showing that irradiation of interstellar ice analogues can produce complex organic residues and prebiotic molecules \citep{MunozCaro2002, Meinert2016}.

\section{Conclusions}

In summary, we have investigated the radiation-driven chemistry of a layered EtA--H$_2$O--CH$_3$OH interstellar ice analogue under 2~keV electron irradiation at 20~K, followed by temperature-programmed desorption up to 310~K. \textit{In situ} FTIR spectroscopy was used to monitor parent-species evolution and irradiation-induced product formation, while \textit{in situ} QMS followed the volatile fraction released during TPD. Complementary \textit{ex situ} ESI-MS analysis provided additional information on the composition of the final residue.

Electron irradiation produces a chemically rich set of products, including CO$_2$, CO, H$_2$CO, H$_2$O$_2$, and nitrogen-bearing species such as NH$_3$, NH$_4^+$-related features, and OCN$^-$. These assignments are consistent with fragmentation, radical recombination, proton-transfer, and acid--base pathways commonly observed in processed water- and methanol-containing molecular ice analogues. The product distribution therefore indicates that the addition of EtA does not suppress the dominant H$_2$O/CH$_3$OH radiolysis chemistry, but introduces additional N-bearing reaction channels.

Signatures consistent with more complex products, including formic acid and polyoxymethylene-like material, are also observed, indicating that carbonyl chemistry and formaldehyde-driven oligomerisation can operate alongside molecular fragmentation in this ternary layered ice. The persistence and sharpening of selected residue features during TPD further suggest that irradiation generates less volatile material whose spectral signatures become more distinct as the ice matrix reorganises during warming.

Comparison with choline chloride--H$_2$O reference spectra supports the tentative association of irradiation-induced bands in the 895--850~cm$^{-1}$ region with methylated N/O-bearing species displaying choline-like spectral signatures. The \textit{ex situ} ESI-MS data likewise show mass-family distributions that resemble the choline chloride reference more closely than the substrate/background blank. In particular, weak higher-$m/z$ families near the nominal mass region of choline-related ions are observed in both the residue and choline chloride reference, whereas no corresponding signals are detected in the blank under the same ESI-MS measurement conditions. However, a secure identification of choline itself cannot be made from the present data because of infrared band overlap, matrix effects, the lack of isolated diagnostic mass-spectral features, and the absence of structural information from the ESI-MS measurements.

The layered geometry is expected to shape the observed chemistry by controlling where the primary electron energy is deposited and where interfacial radical-mediated reactions can occur. The CH$_3$OH and H$_2$O layers lie above the thicker EtA layer and are therefore expected to receive most of the primary electron energy deposition. This is consistent with the dominance of products typical of processed methanol- and water-rich ices, while still allowing chemically relevant EtA processing through direct irradiation of the upper EtA region, interfacial radical chemistry, and secondary-electron-driven reactions.

Further work will be valuable for establishing how broadly the trends identified here apply across different astrochemically relevant ice conditions. In particular, systematic comparisons between layered and fully mixed EtA--H$_2$O--CH$_3$OH ices, together with variations in ice composition and morphology, will help establish how these factors influence the balance between fragmentation, radical recombination, and molecular growth. Experiments spanning a broader range of electron energies and fluences, complemented by more structurally diagnostic analyses of the resulting residues, would further clarify the connection between the reaction pathways identified here and the chemical evolution of irradiated interstellar ices.

Our results show that layered EtA-containing ices are chemically active systems in which fragmentation, radical recombination, and formation of less volatile residue components can occur simultaneously under energetic processing. They support the view that ice composition and layering influence the balance between destruction, preservation, and functionalisation of complex organic molecules, with implications for the evolution of N/O-bearing organic chemistry in irradiated interstellar environments such as the molecular cloud G+0.693$-$0.027 and for the potential delivery of processed organic material to nascent planetary systems.

\section*{Acknowledgements}

This article is based on work from the COST Action CA20129 – Multiscale Irradiation and Chemistry Driven Processes and Related Technologies (MultIChem) and COST Action CA22133 – The Birth of Solar Systems (PLANETS), supported by COST (European Cooperation in Science and Technology). We thank the supplementary financial support from the University of Kent and The Center for Astrochemical Studies at the Max Planck Institute for Extraterrestrial Physics (CAS@MPE). The authors gratefully acknowledge MSc. Diego Enríquez from the Chemical Instrumentation Laboratory (LIQ), Faculty of Engineering, Design and Applied Sciences, Universidad ICESI, for facilitating access to the mass spectrometry facilities and supporting the measurements. We also thank Dr. Gerson Dirceu López Muñoz from the Department of Chemistry, Universidad del Valle, for his helpful discussions and assistance with the interpretation of the mass spectra. The authors acknowledge support from the Europlanet 2024 RI, which has been funded by the European Union’s Horizon 2020 Research Innovation Program under grant agreement no. 871149. The main components of the ICA setup were purchased using funds obtained from the Royal Society through grants UF130409, RGF/EA/180306, and URF/R/191018. Further developments of the installation were supported in part by the Eötvös Loránd Research Network through grants ELKH IF-2/2019 and ELKH IF-5/2020. Support has also been received from the Research, Development, and Innovation Fund of Hungary through grant nos. K128621 and ADVANCED-151196. The research of Z.K. is supported by the Slovak Grant Agency for Science (grant no. 2/0051/26). S.I. acknowledges support from the Danish National Research Foundation through the Center of Excellence “InterCat” (grant agreement no. DNRF150). F.F. acknowledges support from the Marie Skłodowska-Curie Actions (101299799, 101225527), the Royal Society ISPF International Collaboration Awards (ICAO\textbackslash R1\textbackslash 241112, ICAE\textbackslash R1\textbackslash 261037), and the Royal Society of Chemistry Inclusion and Diversity Fund (280026109).

\section*{Data Availability}
The data underlying this work will be shared on reasonable request to the corresponding author.

\bibliographystyle{mnras}
\bibliography{Final_manuscript/MA_mnras}

\clearpage
\appendix
\onecolumn

\section{Infrared bands of aqueous choline chloride solution}

\begin{center}
\includegraphics[width=0.55\textwidth]{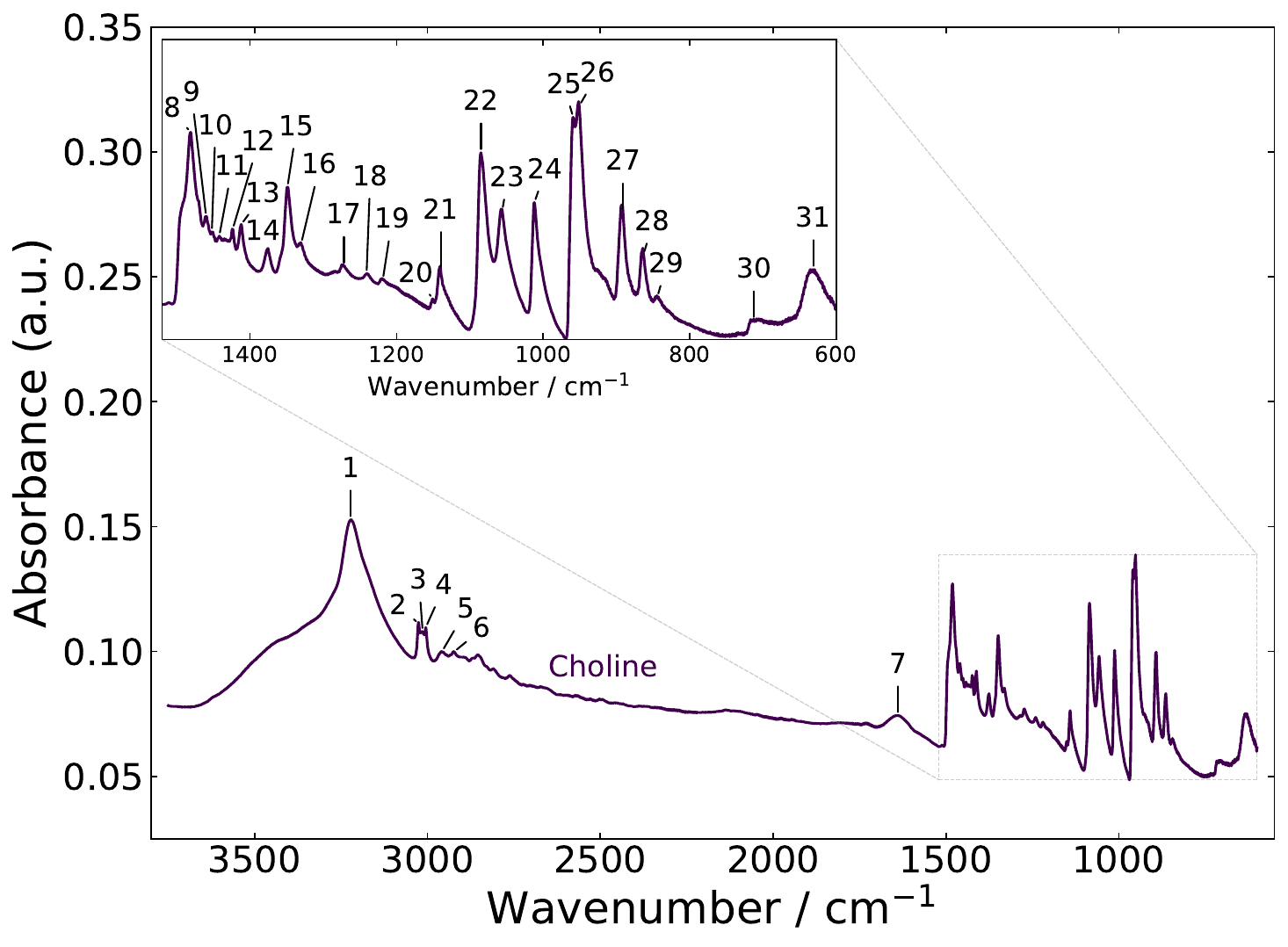}
\captionof{figure}{Mid-infrared spectrum of the choline chloride reference. The inset shows the spectral features between 1520 and 600~cm$^{-1}$ in more detail. The assignment of each band is listed in Table~\ref{tab:Choline-Tab}.}
\label{fig:Choline}
\end{center}

\begin{center}
\captionof{table}{Observed infrared bands and vibrational assignments for the choline chloride reference. The band numbers refer to the band assignments of choline in Fig.~\ref{fig:Choline}.}
\label{tab:Choline-Tab}
\small
\setlength{\tabcolsep}{4pt}
\begin{tabular}{cclccl}
\hline
Band & Wavenumber (cm$^{-1}$) & Mode & Band & Wavenumber (cm$^{-1}$) & Mode\\
\hline
\hline
1 & 3222 & $\nu$(O--H) & 17 & 1272 & $\rho$(CH$_3$) \\
2 & 3025 & $\nu$(CH$_3$), $\nu$(CH$_2$) & 18 & 1241 & $\rho$(CH$_3$) \\
3 & 3014 & $\nu$(CH$_3$), $\nu$(CH$_2$) & 19 & 1219 & $\rho$(CH$_3$) \\
4 & 3005 & $\nu$(CH$_3$), $\nu$(CH$_2$) & 20 & 1151 & $\rho$(CH$_3$) \\
5 & 2958 & $\nu$(CH$_3$), $\nu$(CH$_2$) & 21 & 1140 & $\rho$(CH$_3$) \\
6 & 2923 & $\nu$(CH$_3$), $\nu$(CH$_2$) & 22 & 1085 & $\nu$(C--C), $\nu$(C--O) \\
7 & 1639 & $\delta$(H$_2$O) & 23 & 1056 & $\rho$(CH$_2$) \\
8 & 1481 & $\delta$(CH$_3$) & 24 & 1012 & $\rho$(CH$_2$) \\
9 & 1460 & $\delta$(CH$_3$) & 25 & 959 & $\rho$(CH$_3$), $\nu$(N--CH$_2$) \\
10 & 1452 & $\delta$(CH$_3$) & 26 & 951 & $\nu$(N--CH$_2$) \\
11 & 1442 & $\delta$(CH$_3$), $\delta$\textsubscript{sciss.}(CH$_2$) & 27 & 891 & $\nu$(N--CH$_3$), N(CH$_3$)$_3^{+}$ \\
12 & 1424 & $\delta$(COH) & 28 & 863 & $\nu$(N--CH$_3$), N(CH$_3$)$_3^{+}$ \\
13 & 1412 & $\delta$\textsubscript{sciss.}(CH$_2$) & 29 & 844 & N(CH$_3$)$_3^{+}$ \\
14 & 1375 & $\delta$(CH$_3$) & 30 & 712 & $\nu$(N--CH$_3$), N(CH$_3$)$_3^{+}$ \\
15 & 1349 & $\omega$(CH$_2$) & 31 & 631 & $\tau$(O--H) \\
16 & 1330 & $\omega$(CH$_2$) & & & \\
\hline
\hline
\end{tabular}

\par\vspace{0.4em}
\begin{minipage}{0.95\textwidth}
\footnotesize
\raggedright
\textbf{Modes:} $\nu$ (stretching), $\delta$ (bending), $\omega$ (umbrella), $\rho$ (rocking), $\tau$ (twisting).\par
\textbf{Assignments:} Based on previous studies by \citet{millerAqueousInfraredPharmaceutical1988,pawlukojcINSDFTTemperature2014,golestanifarIntroductionCharacterizationNovel2025}.\par
\vspace{0.3em}
The choline chloride reference was prepared from a 0.1~M aqueous choline chloride solution, which was drop-cast onto the ZnSe substrate before introduction into the chamber at room temperature, approximately 310~K.
\end{minipage}
\end{center}

\bsp
\label{lastpage}
\end{document}